\documentclass[journal]{IEEEtran}
\IEEEoverridecommandlockouts
\usepackage{cite}
\usepackage{pifont}
\usepackage{url}
\usepackage{hyperref}
\usepackage{subcaption}
\usepackage{textcase}
\usepackage{multicol,multirow}
\usepackage[utf8]{inputenc}
\usepackage{pgfplots}
\usepgfplotslibrary{groupplots,dateplot}
\usetikzlibrary{patterns,shapes.arrows}
\pgfplotsset{compat=newest}
\usepackage{amsmath,amssymb,amsfonts}
\usepackage{algorithmic}
\usepackage{graphicx}
\usepackage{diagbox}
\usepackage{textcomp}
\usepackage{xcolor}
\usepackage{enumitem}
\usepackage{mathtools}
\usepackage{mathstuffantoine}
\newtheorem{assumption}{Assumption}
\def\BibTeX{{\rm B\kern-.05em{\sc i\kern-.025em b}\kern-.08em
    T\kern-.1667em\lower.7ex\hbox{E}\kern-.125emX}}

\newcommand\copyrighttext{%
  \footnotesize \textcopyright 2024 IEEE. Personal use of this material is permitted.
  Permission from IEEE must be obtained for all other uses, in any current or future 
  media, including reprinting/republishing this material for advertising or promotional 
  purposes, creating new collective works, for resale or redistribution to servers or 
  lists, or reuse of any copyrighted component of this work in other works. 
  DOI: \href{10.1109/TSP.2024.3383277}{10.1109/TSP.2024.3383277}}
\newcommand\copyrightnotice{%
\begin{tikzpicture}[remember picture,overlay]
\node[anchor=south,yshift=5pt] at (current page.south) {\fbox{\parbox{\dimexpr\textwidth-\fboxsep-\fboxrule\relax}{\copyrighttext}}};
\end{tikzpicture}%
}

\begin{document}

\title{
DANSE: Data-driven Non-linear State Estimation of Model-free Process in Unsupervised Learning Setup 
}

\author{
    \IEEEauthorblockN{Anubhab Ghosh, Antoine Honoré, Saikat Chatterjee} 
    
    \IEEEauthorblockA{Digital Futures Centre, and School of Electrical Engineering and Computer Science \\ KTH Royal Institute of Technology, Stockholm, Sweden\\
    anubhabg@kth.se, honore@kth.se, sach@kth.se}

    \thanks{
    The work is supported by a research grant from Digital Futures Centre, KTH, Sweden. Project title: ``DLL: Data-limited learning of complex systems''. Website: https://www.digitalfutures.kth.se/.
    
    Reproducible code is available: https://github.com/anubhabghosh/danse\_jrnl and https://github.com/saikatchatt/danse-jrnl.}
}

\maketitle
\copyrightnotice

\begin{abstract}
We address the tasks of Bayesian state estimation and forecasting for a model-free process in an unsupervised learning setup. {For a model-free process, we do not have any a-priori knowledge of the process dynamics.} In the article, we propose DANSE -- a \emph{Da}ta-driven \emph{N}onlinear \emph{S}tate \emph{E}stimation method. DANSE provides a closed-form posterior of the state of the model-free process, given linear measurements of the state. 
In addition, it provides a closed-form posterior for forecasting. A data-driven recurrent neural network (RNN) is used in DANSE to provide the parameters of a prior of the state. The prior depends on the past measurements as input, and then we find the closed-form posterior of the state using the current measurement as input. The data-driven RNN captures the underlying non-linear dynamics of the model-free process. The training of DANSE, mainly learning the parameters of the RNN, is executed using an unsupervised learning approach. In unsupervised learning, we have access to a training dataset comprising only a set of (noisy) measurement data trajectories, but we do not have any access to the state trajectories. Therefore, DANSE does not have access to state information in the training data and can not use supervised learning. Using simulated linear and non-linear process models (Lorenz attractor and Chen attractor), we evaluate the unsupervised learning-based DANSE. We show that the proposed DANSE, without knowledge of the process model and without supervised learning, provides a competitive performance against model-driven methods, such as the Kalman filter (KF), extended KF (EKF), unscented KF (UKF), {a data-driven deep Markov model (DMM)} and a recently proposed hybrid method called KalmanNet. {In addition, we show that DANSE works for high-dimensional state estimation.} 

\end{abstract}

\begin{IEEEkeywords}
Bayesian state estimation, forecasting, neural networks, recurrent neural networks, unsupervised learning.
\end{IEEEkeywords}

\section{Introduction}\label{sec:introduction}
\noindent\emph{Context:} Estimating the state $\mathbf{x}_t$ of a complex dynamical system/process from noisy measurements/observations $\mathbf{y}_t$, maintaining causality, is a fundamental problem in the field of signal processing, control, and machine learning. Here, $t$ denotes a discrete time index. {The Bayesian state estimation task is to find a closed-form posterior of the state given the current and past measurements, as $p(\mathbf{x}_t \vert \mathbf{y}_t, \mathbf{y}_{t-1}, \hdots, \mathbf{y}_1)$.}

For the state estimation problem, traditional Bayesian methods assume that the model of state dynamics (also called process model or state space model) is known. Examples of such model-driven methods are the Kalman filter (KF) and its extensions such as the extended KF (EKF), the unscented KF (UKF), and sampling-based methods such as particle filters \cite{kalman1961new, kalman1960new, gruber1967approach, julier2004unscented, arulampalam2002tutorial}. We assume that the state space model (SSM) in a model-driven method (also referred to as model-based) is formed using the physics of the process and/or some a-priori knowledge, and there is no need for learning (or training). On the other hand, modern data-driven methods could use deep neural networks (DNNs), for example, recurrent neural networks (RNNs), to map $\mathbf{y}_t$ to $\mathbf{x}_t$ directly, without knowing the process model. The data-driven methods can handle complex model-free processes using a supervised learning-based training approach. At the training phase, supervised learning requires to access state $\mathbf{x}_t$ to have a labelled dataset in a pairwise formation $(\mathbf{y}_t,\mathbf{x}_t)$. The labelled dataset is mainly used to train data-driven RNNs.

\vspace{2mm}
\noindent\emph{Motivation:} There are many scenarios where the physics of the process is unknown or limited, and the state $\mathbf{x}_t$ is not accessible due to practical reasons. Inaccessible state $\mathbf{x}_t$ does not allow to have a labelled dataset. 
{We have to deal with a model-free process. Note that, for a model-free process we do not have any a-priori knowledge of the process, for example, whether the process is Markovian; we do not know the SSM, the structure of the SSM, and the SSM's parameters.} Therefore, we can not engineer model-driven methods and supervised learning-based data-driven methods. 

\vspace{2mm}
\noindent\emph{{Research question:}}
Can we develop a Bayesian estimator of $\mathbf{x}_t$ for a complex model-free process without any access to  $\mathbf{x}_t$, but using an unsupervised learning approach based on a training dataset comprised of noisy measurements {$\lbrace  \mathbf{y}_{\tau} \rbrace$}? 
The \emph{purpose} of the article is to develop DANSE (\emph{Da}ta-driven \emph{N}onlinear \emph{S}tate \emph{E}stimation) method to answer the above question, based on our recent contribution \cite{ghosh2023danse}.



\vspace{2mm}
\noindent\emph{Potential applications:} DANSE has potential applications when there are enough measurements for unsupervised learning, but it is difficult to access the state for supervised learning and/or design a good tractable process model. For example, we can explore the use of DANSE for internal state estimation of a medical patient from physiological time-series data. Another example is positioning and navigation of autonomous systems in hazardous, underwater, off-road conditions where GPS (global positioning system) signal is not available to know the state \cite{guo2018vehicle} and create labelled data for supervised learning.

\vspace{2mm}
\noindent\emph{Aspects of DANSE:} 
\begin{enumerate}[noitemsep,nolistsep]
\item It is for a \emph{model-free} process where the state dynamics are expected to be non-linear and complex.
\item \emph{Unsupervised learning} of DANSE is realized using a standard maximum-likelihood principle. 
\item For linear measurements, it provides \emph{closed-form} posterior $p(\mathbf{x}_t|\mathbf{y}_t,\mathbf{y}_{t-1}, \hdots, \mathbf{y}_1)$ in terms of RNN parameters.
\item It is a causal system, and suitable for \emph{forecasting}. 
\end{enumerate}

\begin{table*}[htbp]
    \centering
    \scalebox{1.2}{\begin{tabular}{cccccc}  
        \hline
        \multicolumn{1}{c}{Attribute} &
        \multicolumn{5}{c}{Methods/Approaches} \\\hline
        \multicolumn{1}{c}{} &
        \multicolumn{4}{c}{Existing Methods/Approaches} &
        \multicolumn{1}{c}{\textbf{DANSE}}
        \\ \cline{2-5}
        {} & Model-based & \begin{tabular}{@{}c@{}}{Data-driven}\\ {Non-parametric}\end{tabular} & \begin{tabular}{@{}c@{}}{Data-driven}\\ {Parametric}\end{tabular} & Hybrid &   \\
         & e.g. KF, EKF, etc. & e.g. {GPs, etc.} & e.g. DVAE, RNNs, etc.  & e.g. KalmanNet & \\
        \hline\hline
        \multicolumn{5}{c}{Domain knowledge} \\\hline
        Process state dynamics (SSM) & \ding{51} & {\ding{55}} & \ding{55} & \ding{51} & \ding{55} \\
        Process noise statistics & \ding{51} & {\ding{55}} & \ding{55} & \ding{55} & \ding{55}\\
        \hline
        {Measurement system} & \ding{51} & {\ding{55}} & \ding{55} & \ding{51} & \ding{51} \\
        Measurement noise statistics & \ding{51} & {\ding{55}} & \ding{55} & \ding{55} & \ding{51}\\
        \hline\hline
        \multicolumn{5}{c}{Learning and Inference} \\\hline
        Supervised learning & --- & {\ding{51} / \ding{55}} & \ding{51} / \ding{55}  & \ding{51} / \ding{55} & \ding{55} \\\hline
        Closed-form posterior & \ding{51} & {\ding{51} / \ding{55}} & \ding{55} & \ding{51} & \ding{51} \\
        \hline
    \end{tabular}}
    \caption{A visualization of attributes for different existing methods/approaches -- model-driven, data-driven, hybrid, -- and the proposed DANSE. The symbol `---' indicates the non-applicability of the method/approach to the attribute, `\ding{51}' indicates that the aspect is required/satisfied, and `\ding{55}' indicates that the aspect is not required/satisfied. 
    }
    \label{tab:danse_concept}
\end{table*} 
\subsection{Literature Survey}\label{sec:literature_survey}

There can be three broad approaches for state estimation: a model-based approach, a data-driven approach, or a combination of the two. 

\noindent\textit{Model-based approach}: The model-based approach is traditional, and the famous example is KF \cite{kalman1960new, kalman1961new}. In KF,  we use a linear SSM and there is an explicit measurement {system} which is also linear \cite{kalman1960new, kalman1961new}. In the linear SSM, the current state $\mathbf{x}_t$ depends linearly on the previous state $\mathbf{x}_{t-1}$ and driving process noise. Using Gaussian distributions for all relevant variables as a-priori statistical knowledge and the Markovian relation between states that follow the linear model, KF is analytically tractable. KF is known to be an optimal Bayesian estimator that achieves minimum-mean-square-error (MMSE). 

Since dynamical systems are often non-linear in nature, an extension of the Kalman filter, namely the EKF, was proposed to account for the non-linear variations \cite{gruber1967approach}. The beauty of the EKF scheme is that it maintains analytical tractability like KF using a linear, time-varying approximation of non-linear functions. This is engineered using a first-order Taylor series approximation of the underlying state space model and measurement system, that is, computing the respective Jacobians. An extension to second-order extended Kalman filters was proposed in \cite{athans1968suboptimal}. The second-order approach aims to reduce the error caused by simple linearization around the current estimate using a second-order Taylor series expansion. 

Another non-linear extension of the KF is the unscented Kalman filter (UKF) that seeks to approximate the SSM using a derivative-free approximation as opposed to EKF \cite{julier2004unscented, wan2000unscented}. The main idea of UKF is using a set of specific points called \textit{sigma points} to approximate the probability density function of an unknown distribution. Instead of propagating state estimates and covariances, the algorithm propagates these sigma points that are subsequently weighted-averaged to yield the required state estimates and covariances. Another approach involving sigma-points includes the cubature KF \cite{arasaratnam2009cubature}, which enables applying such methods to both low and high-dimensional problems. 

It is worthwhile to note that the above-mentioned schemes -- KF, EKF, and UKF are commonly used with additive, Gaussian noise dynamics in the state-space models, although there are means to extend them to the case of more general noise dynamics as well \cite[Chap. 13, 14]{simon2006optimal}. To handle the numerical issues associated with the propagation of errors of the state covariance matrices, square-root algorithms have also been developed for KF, EKF, and UKF in \cite{kaminski1971discrete, park1995new, van2001square}.  
Schemes based on sequential Monte Carlo (SMC) sampling such as the particle filters (PFs) are also capable of handling non-linear, non-Gaussian dynamics but are often computationally intensive \cite{doucet2001sequential, arulampalam2002tutorial}. A review of model-based approaches has been provided in \cite{patwardhan2012nonlinear}. 

\noindent\textit{Data-driven approach}: The second broad approach is the data-driven one. {In this case, there are two broad directions -- non-parametric and parametric approaches. Non-parametric approaches involve using flexible models such as Gaussian processes (GPs) for modeling the state and measurement dynamics. Gaussian processes (GP) are flexible, non-parametric models that can be either integrated into various parts of common model-based filters such as KF, EKF, UKF or PF \cite{ko2009gp} or can be even used for target tracking in a recursive, online manner \cite{aftab2020learning}. Typically the learnable parameters of the GPs include hyperparameters for the mean and covariance functions which are often learned in a supervised manner as in \cite{ko2009gp} or in an unsupervised manner \cite{aftab2020learning}. Often, the computational complexity of such a method becomes prohibitively high for longer sequences, thus requiring certain numerical approximations to be made. This can include using a fixed number of sigma points (as in a UKF) for approximating probability distributions, or a smaller context window of inducing points for modeling longer sequences \cite{aftab2020learning}, usually one sequence at a time}. 

Generally, a data-driven approach does not require explicit models, such as the Markovian relation between states. Also, they do not utilize an explicit measurement system. Approaches for state estimation using prediction error methods and RNNs were proposed in \cite{yadaiah2006neural}. Approximation of the unknown process and measurement noises at each time step using DNNs with measurement data inputs was explored in \cite{xu2021ekfnet}. {The use of either fixed-order deterministic functions such as polynomial of basis functions was investigated in \cite{li2018joint, li2023target}. Lastly, parametric approaches also include the use of Deep neural networks (DNNs), mainly recurrent neural networks (RNNs) such as gated recurrent units (GRUs) \cite{choPropertiesNeuralMachine2014} and long short-term memory networks (LSTMs) \cite{hochreiter1997long}.} 
The use of RNNs such as LSTMs for learning the parameters of the underlying SSM in an end-to-end, supervised manner was proposed in \cite{coskun2017long}. The use of LSTMs alleviated the need for linear transitions in the state, but in turn, the method required the knowledge of the true state during training. That means it uses supervised learning with labelled data.

Another set of approaches involves modeling the distributions of the underlying SSM using (deep) neural networks. 
Example schemes include the class of dynamical variational autoencoders (DVAEs) \cite{girin2021dynamical, fraccaro2017disentangled, krishnan2017structured} where training is performed using approximate Bayesian inference called variational inference (VI). In \cite{fraccaro2017disentangled}, a variational autoencoder was used in combination with a linear Gaussian SSM, leading to the Kalman variational autoencoder (KVAE), and the overall model was trained in an unsupervised manner. Another VI-based approach known as deep variational Bayes filters (DVBFs) performs state estimation by inferring the process noise from the observations \cite{karl2017deep} while assuming the state obeys locally-linear transitions. Another approach uses ideas similar to kernel-based approaches for learning a high-dimensional state \cite{becker2019recurrent}. The use of VI and the complex relational structures have the disadvantages of (1) not providing closed-form Bayesian posterior of the state, owing to the black-box nature of purely data-driven methods, and (2) leading to computationally intensive training, {involving a cascade of expectations and approximating each expectation by using Monte-Carlo approximations and the re-parameterization trick \cite{kingma2013auto}}. 
{Additionally in case of DVAE models such as deep Markov models (DMMs) \cite{krishnan2017structured}, the approximate posterior for the state $\mathbf{x}_t$ at time $t$ relies on the previous state $\mathbf{x}_{t-1}$ as well as future noisy measurements $\mathbf{y}_{1:\tau}$ with $\tau > t$, and is thus inherently non-causal.} {While it is possible to employ causal inference schemes, the Markovian structure of the underlying state is inherent in DVAE models. The optimization methods for learning require sequential sampling processes like ancestral sampling. Hence, the learning procedure of DVAE methods is computationally intensive and prone to approximation errors. In addition, the DVAE methods are mainly designed for modeling observed sequential data and are not directly designed to address the (causal) Bayesian state estimation task using a known measurement system. Bayesian state estimation can be visualized as a byproduct.} 

{Finally,} if we have access to both measurements $\mathbf{y}_{1:t}$ and states $\mathbf{x}_{1:t}$ as training data, then we can directly use RNNs for state estimation by minimizing a loss based on $\mathbf{y}_{1:t}$ and $\mathbf{x}_{1:t}$. This is a supervised training approach using labeled data. We can also use sequence-to-sequence modeling methods, such as Transformers \cite{vaswani2017attention}. The problem is that in most of the cases, we do not have access to states $\mathbf{x}_{1:t}$, and hence such direct use of RNNs and Transformers is difficult to realize. Furthermore, they typically yield a point estimate and do not provide posterior distributions. {There also exist specific data-driven approaches for target tracking, e.g. the use of recurrent neural networks for modeling the state transition dynamics of an unscented Kalman filter as in \cite{li2021recurrent} or the use of an LSTM to model a time-varying Kalman filter as proposed in \cite{song2022improved}.}

\noindent\textit{Hybrid approach}: The third broad approach, known as the hybrid approach, seeks to use the best of both worlds. It endeavors to amalgamate both model-driven and data-driven approaches. A recent example is the KalmanNet method \cite{revach2022kalmannet} that proposes an online, recursive, low-complexity, and data-efficient scheme based on the KF architecture. KalmanNet involves modeling the Kalman gain using DNNs, thus maintaining the structure of the model-based KF while incorporating some data-driven aspects. The method uses a supervised learning approach where we need labelled data comprised of true states and noisy observations. However, it is shown robust to the noise statistics. Experimentally, it is also shown to perform well in cases when partial information about SSM is available or in cases of model mismatch. A modification to the above scheme is also proposed as an unsupervised KalmanNet in \cite{revach2022unsupervised}, which uses only noisy measurements as training data. The KalmanNet idea is further extended to other tasks such as smoothing {using RTSNet} \cite{ni2022rtsnet} {or specifically for handling partial model information or model mismatch using Split-KalmanNet \cite{choi2023split}. A common attribute of the data-driven and hybrid approaches is that analytically obtaining statistical properties for such estimators is quite challenging.} 

A comparison between model-based, data-driven, hybrid methods/approaches and DANSE is shown in Table \ref{tab:danse_concept}. 
\subsection{{Contributions}}\label{sec:contributions}
{ In this contribution, we build upon our previous work \cite{ghosh2023danse} and extend it in the following manner:
\begin{enumerate}[noitemsep,nolistsep]
    \item We illustrate the performance of DANSE for nonlinear state estimation tasks vis-á-vis model-based, {data-driven} and hybrid approaches through further extensive experiments on nonlinear models such as the Lorenz attractors \cite{lorenz1963deterministic}. 
    \item We propose an empirical estimation performance limit using a supervised learning scenario to compare the performance of DANSE and other state estimation approaches on nonlinear state estimation tasks. 
    \item We empirically demonstrate the effect of training data on the performance of DANSE, robustness to mismatched noise conditions, and its ability to perform one-step ahead forecasting and state prediction. The advantage of DANSE is that the learning problem formulation accommodates these two tasks besides state estimation.
    \item Finally, we demonstrate the ability of DANSE to track high-dimensional states for a nonlinear dynamical system such as a 20-dimensional Lorenz-96 attractor \cite{lorenz1996predictability, lorenz1998optimal}.
\end{enumerate}
}
\subsection{Notations}\label{sec:notations}
We use bold font lowercase symbols to denote vectors and regular lowercase font to denote scalars, for example, $\mathbf{x}$ represents a vector while $x_{j}$ represents the $j$'th component of $\mathbf{x}$. A sequence of vectors $\mathbf{x}_1, \mathbf{x}_2, \ldots, \mathbf{x}_t$ is compactly denoted by $\mathbf{x}_{1:t}$, where $t$ denotes a discrete time index. Then $x_{t,j}$ denotes the $j$'th component of $\mathbf{x}_t$. {In the case of continuous-time indices, the distinction is mentioned clearly using a slight abuse of notation}. Upper case symbols in bold font, like $\mathbf{H}$, represent matrices. The operator $\left(\cdot\right)^{\top}$ denotes the transpose. $\mathcal{N}\left(\cdot; \mathbf{m}, \mathbf{L} \right)$ represents the probability density function of the Gaussian distribution with mean $\mathbf{m}$ and covariance matrix $\mathbf{L}$. $\mathbb{E}\lbrace \cdot \rbrace$ denotes the expectation operator. The notation $\Vert \mathbf{x} \Vert^2_{\mathbf{C}}$ denotes the squared $\ell_2$ norm of $\mathbf{x}$ weighted by the matrix $\mathbf{C}$, i.e. $\Vert \mathbf{x} \Vert^2_{\mathbf{C}} = \mathbf{x}^{\top} \mathbf{C} \mathbf{x}$.

\subsection{Structure of the article}\label{sec:strcuture}
The article is organized as follows: in section \ref{sec:proposed_danse}, we describe an unsupervised learning setup and the proposed DANSE in detail, including a proposed empirical estimation performance limit. Next, we show experiments in section \ref{sec:experiments_and_results}. We demonstrate the empirical performance of DANSE for a linear SSM as a proof-of-concept, and then for three nonlinear SSMs - the Lorenz, Chen and Lorenz-96 attractors \cite{lorenz1963deterministic, chen1999yet, lorenz1996predictability}. We experimentally compare DANSE vis-\`a-vis KF (for the linear SSM), EKF, UKF, {DMM} \cite{krishnan2017structured} and the unsupervised KalmanNet \cite{revach2022unsupervised} (for the nonlinear SSMs). Additionally, we endeavor to empirically answer pertinent questions regarding training data requirements and robustness to mismatched conditions during inference. Finally, we provide conclusions and the scope of future works in section \ref{sec:conclusions}. 

\section{Proposed DANSE} \label{sec:proposed_danse}
\subsection{Bayesian Inference and Unsupervised Learning Setup}
\label{subsec:SystemSetup}
Let there be a dynamical signal $\mathbf{x}_{1:T}$ (with $\mathbf{x}_t \in \mathbb{R}^m$), representing a model-free process of length $T$. The process is expected to be complex and we have no prior knowledge of the process. Neither do we know its statistical properties nor have direct access to the process data in the learning stage. We assume that we have access to the linear measurements $\mathbf{y}_t \in \mathbb{R}^n$ of the process, where
\begin{equation}
\label{eq:measurement}
\mathbf{y}_t = \mathbf{H}_t \mathbf{x}_t + \mathbf{w}_t, \,\, t=1,2, \hdots, T.
\end{equation}
Here $\mathbf{w}_t \sim \mathcal{N}\left(\boldsymbol{0}, \mathbf{C}_w\right)$ is a standard measurement noise with zero mean and covariance $\mathbf{C}_w \in \mathbb{R}^{n \times n}$, and $\mathbf{H}_t \in \mathbb{R}^{n \times m}$ denotes the known measurement {matrix. Together, they constitute the linear measurement system in \eqref{eq:measurement}.} 
Maintaining causality, the Bayesian inference tasks are mentioned below. 

\begin{enumerate}
    \item[(T1)] \emph{State estimation problem:} The inference task is to estimate the posterior of the current state $\mathbf{x}_t$ given $\mathbf{y}_{1:t}$, denoted by $p(\mathbf{x}_t|\mathbf{y}_1,\mathbf{y}_2, \hdots, \mathbf{y}_t) 
     \triangleq p(\mathbf{x}_t|\mathbf{y}_{1:t})$. In addition to estimate the posterior of the time series $\mathbf{x}_{1:t}$, denoted by $p(\mathbf{x}_1,\mathbf{x}_2, \hdots, \mathbf{x}_t|\mathbf{y}_1,\mathbf{y}_2, \hdots, \mathbf{y}_t) \triangleq p(\mathbf{x}_{1:t}|\mathbf{y}_{1:t})$. 
     \item[(T2)] \emph{Forecasting problem:} The inference task is to estimate the distribution of the future measurement $\mathbf{y}_{t+1}$ given $\mathbf{y}_{1:t}$, denoted by $p(\mathbf{y}_{t+1}|\mathbf{y}_{1:t})$, and also optionally that of the future state $\mathbf{x}_{t+1}$ given $\mathbf{y}_{1:t}$, denoted by $p(\mathbf{x}_{t+1}|\mathbf{y}_{1:t})$.  
\end{enumerate}
\begin{figure}[t]
    \centering
    \begin{minipage}{0.5\textwidth}
    \centering
    \includegraphics[width=1\textwidth]{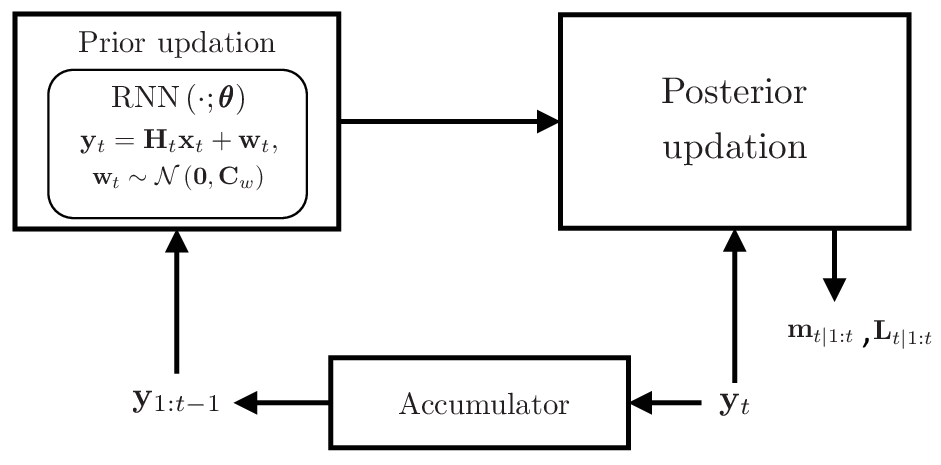}
    \subcaption{}
    \end{minipage}  
    \vspace{0.1 in}
    \begin{minipage}{0.5\textwidth}
    \centering
    \includegraphics[width=1\textwidth]{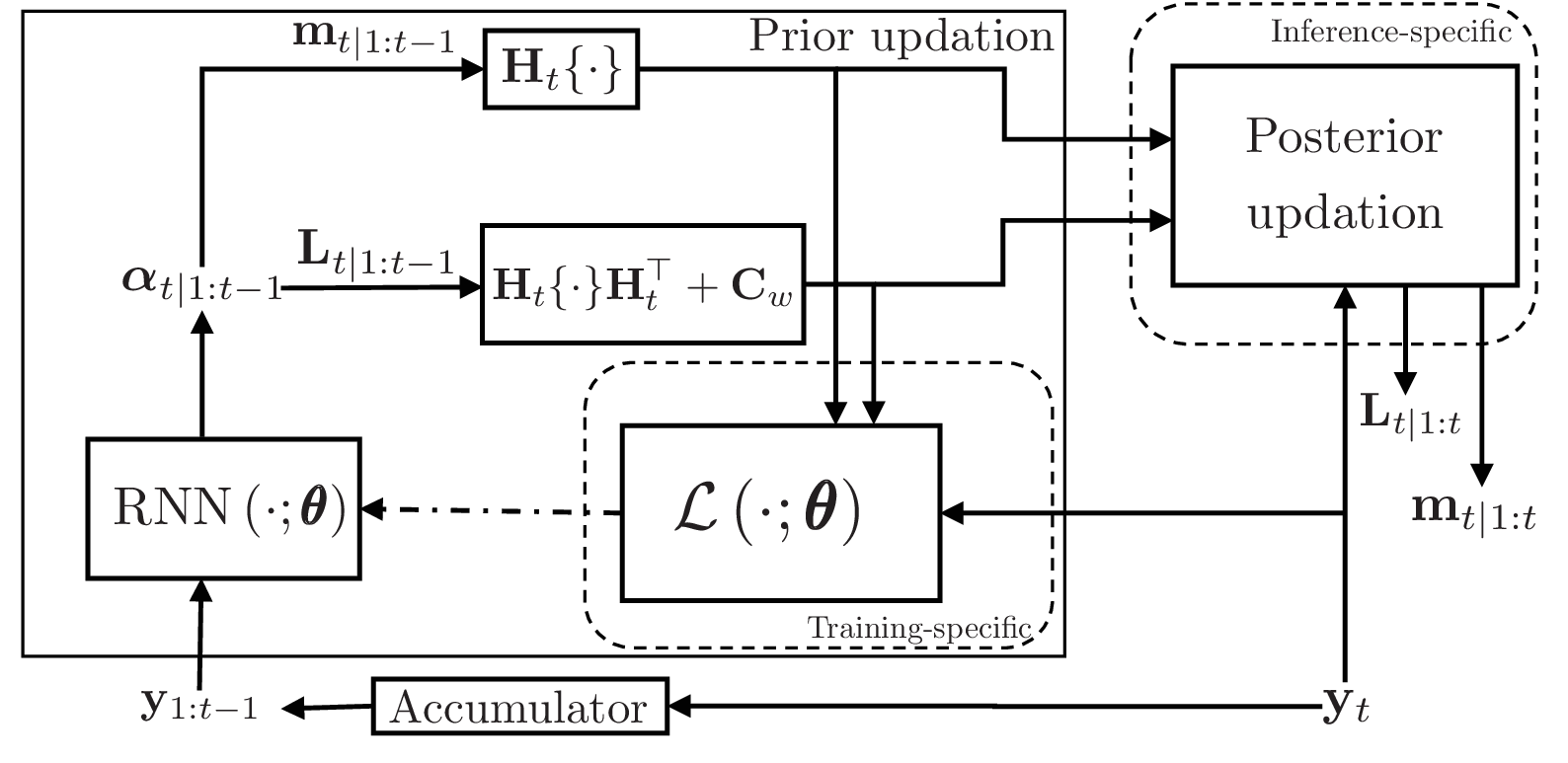}
    \subcaption{}
    \end{minipage}
    \caption{DANSE architecture. (a) Simple block diagram for inference. (b) Detailed diagram. The dash-dotted line represents the gradient flow during the training phase, solid lines indicate information flow during training/inference. Training/inference-specific blocks are shown within dashed borders. The `Accumulator' block collects observations recursively.}
    \label{fig:danse}
    
\end{figure}

{To learn the parameters of DANSE, we have a training dataset $\mathcal{D}$ comprised of $N$ time-series measurements as 
\begin{eqnarray}\label{eq:dataset_Def}
 \mathcal{D} = \left\{ \mathbf{y}^{(i)}_{1:T^{(i)}} \right\}_{i=1}^N.   
\end{eqnarray}
}
Here $ \mathbf{y}^{(i)}_{1:T^{(i)}} = \mathbf{y}_1^{(i)},\mathbf{y}_2^{(i)}, \hdots, \mathbf{y}_{T^{(i)}}^{(i)} $ is the $i'\text{th}$ time-series measurements of length $T^{(i)}$. Note that $T^{(i)}$ can vary across samples in $\mathcal{D}$; that means, the time-series measurements can have unequal lengths. Since we do not have access to the corresponding state sequences, $\mathcal{D}$ is unlabelled, and the learning problem requires to be unsupervised.  

\subsection{DANSE System}\label{sec:danse_system}
DANSE is model-free. It only has access to $\mathcal{D}$. DANSE is data-driven and seeks to have analytical tractability like model-driven methods. So, to use data and the measurement system together, a prior-posterior system setup needs to be developed where the prior comes from data and the posterior follows the constraint of the linear measurement system. 
Principally, we model the unknown prior probability distribution $p\left(\mathbf{x}_t \vert \mathbf{y}_{1:t-1} \right)$ as a Gaussian distribution parameterized by an RNN. 

At time instant $t$, an RNN recursively uses the input sequence $\mathbf{y}_{1:t-1}$ and provides the parameters of the Gaussian prior $p\left(\mathbf{x}_t \vert \mathbf{y}_{1:t-1}\right)$ as the output, collectively denoted as $\boldsymbol{\alpha}_{t \vert 1:t-1}$. An RNN has its parameters $\pmb{\theta}$, thus its output also depends on $\pmb{\theta}$. To indicate this, we write $\boldsymbol{\alpha}_{t \vert 1:t-1} \triangleq \boldsymbol{\alpha}_{t \vert 1:t-1}(\pmb{\theta})$.  
{Then, at the time instant $t$, DANSE computes the posterior $p\left(\mathbf{x}_t \vert \mathbf{y}_{1:t}\right)$ based on the measurement system \eqref{eq:measurement} and the RNN-based prior $p(\mathbf{x}_{t}|\mathbf{y}_{1:t-1})$}. 
A simple block diagram of the Bayesian estimation as the inference of DANSE is shown in Fig. \ref{fig:danse}(a). 

\noindent{\emph{Bayesian State Estimation (Solution of T1)}:}
The task is to obtain current state posterior $p(\mathbf{x}_t|\mathbf{y}_1, \ldots, \mathbf{y}_{t-1},\mathbf{y}_t) = p(\mathbf{x}_t|\mathbf{y}_{1:t})$, and the time-series posterior $p(\mathbf{x}_{1:t}|\mathbf{y}_{1:t})$. The prior distribution $p\left(\mathbf{x}_t \vert \mathbf{y}_{1:t-1}\right)$ and the observation distribution $p\left(\mathbf{y}_t \vert \mathbf{x}_t \right)$ are shown below in \eqref{eq:prior_lik}. 
\begin{eqnarray}
\label{eq:prior_lik}
\begin{array}{c}
\mathrm{Prior:} \, p(\mathbf{x}_t|  \mathbf{y}_{1:t-1}) \!=\! \mathcal{N}(\mathbf{x}_t; \mathbf{m}_{t|1:t-1}(\pmb{\theta}),  \mathbf{L}_{t|1:t-1}(\pmb{\theta}) ), \\
\text{ such that } \lbrace \mathbf{m}_{t|1:t-1}(\pmb{\theta}), \mathbf{L}_{t|1:t-1}(\pmb{\theta}) \rbrace \triangleq \boldsymbol{\alpha}_{t \vert 1:t-1}(\pmb{\theta}), \\
\boldsymbol{\alpha}_{t \vert 1:t-1} \left(\pmb{\theta}\right) = \mathrm{RNN} (\mathbf{y}_{1:t-1}; \pmb{\theta}). \\
\mathrm{Observation:} \, p(\mathbf{y}_t | \mathbf{x}_t ) = \mathcal{N}(\mathbf{y}_t ; \mathbf{H}_t\mathbf{x}_t, \mathbf{C}_w ).
\end{array}
\end{eqnarray}
Here, $\mathbf{m}_{t|1:t-1}(\pmb{\theta}) \in \mathbb{R}^m$ and $\mathbf{L}_{t|1:t-1}(\pmb{\theta}) \in \mathbb{R}^{m \times m}$ denote the mean and covariance matrix of the Gaussian prior distribution, respectively. The mean and covariance depend on RNN parameters $\pmb{\theta}$, and hence we show it in notations. We use $\mathbf{L}_{t|1:t-1}(\pmb{\theta})$ as a diagonal covariance matrix for ease of practical implementation. 
Then, using the `completing the square' approach \cite[Chap. 2]{bishop2006pattern}, the posterior distribution of the current state  $p\left(\mathbf{x}_t \vert \mathbf{y}_{1:t}\right)$ is obtained in closed-form as 
\begin{eqnarray}
\label{eq:posterior_update}
\begin{array}{rl}
&p(\mathbf{x}_t|\mathbf{y}_{1:t}) =  \mathcal{N}( \mathbf{x}_t ; \mathbf{m}_{t|1:t}(\pmb{\theta}), \mathbf{L}_{t|1:t}(\pmb{\theta})), \\
&\mathbf{m}_{t|1:t}(\pmb{\theta}) 
= \mathbf{m}_{t|1:t-1}(\pmb{\theta}) + \mathbf{K}_{t\vert 1:t-1} \pmb{\varepsilon}_t, \\
&\mathbf{L}_{t|1:t}(\pmb{\theta}) 
= \mathbf{L}_{t|1:t-1}(\pmb{\theta}) -  \mathbf{K}_{t\vert 1:t-1}\mathbf{R}_{\varepsilon} \mathbf{K}_{t\vert 1:t-1}^{\top},\\
\end{array}
\end{eqnarray}
where the second equation in \eqref{eq:posterior_update} is obtained using the Woodbury matrix identity/matrix inversion lemma, with 
\begin{equation}\label{eq:posterior_update_terms}
    \begin{array}{l}
    \mathbf{K}_{t\vert 1:t-1}  \triangleq \mathbf{L}_{t|1:t-1}(\pmb{\theta})\mathbf{H}_t^{\top}\mathbf{R}_{\varepsilon}^{-1}, \\ \mathbf{R}_{\varepsilon}  \triangleq \mathbf{H}_t\mathbf{L}_{t|1:t-1}\left(\pmb{\theta}\right)\mathbf{H}_t^{\top} + \mathbf{C}_w, \,\, \mathrm{and} \\
    {\pmb{\varepsilon}}_t  \triangleq \mathbf{y}_t - \mathbf{H}_t\mathbf{m}_{t|1:t-1}(\pmb{\theta}). 
    \end{array}
\end{equation}
A point estimate can be $\mathbf{m}_{t|1:t}({\pmb{\theta}})$, that provides a practical mean-square-error (MSE) performance as $\mathbb{E}\{ \| \mathbf{x}_t - \mathbf{m}_{t|1:t}(\pmb{\theta}) \|_2^2\}$. 
{Finally, we are interested in the closed form posterior $p\left(\mathbf{x}_{1:t} \vert \mathbf{y}_{1:t} \right)$. We use the following assumption.
\begin{assumption}\label{assum:assumption1}
    We have a causal filtering scenario, where we neither have information regarding the SSM nor access to the true states $\mathbf{x}_t$ during inference. Hence we assume 
\begin{equation}\label{eq:posterior_assumption}
    p\left(\mathbf{x}_t \vert \mathbf{y}_{1:t}, \mathbf{x}_{1:t-1}\right) = p\left(\mathbf{x}_t \vert \mathbf{y}_{1:t}\right) 
\end{equation}
\end{assumption}
}
{Using  \eqref{eq:posterior_update} and assumption \ref{assum:assumption1}, we write the joint posterior as}
\begin{eqnarray}
\label{eq:posterior_at_time_t}
\begin{array}{rl}
    p\!\left(\mathbf{x}_{1:t} \vert \mathbf{y}_{1:t}\right) \!\!\!\!\!\!&= p\left(\mathbf{x}_t \vert \mathbf{y}_{1:t}, \mathbf{x}_{1:t-1}\right) p\left(\mathbf{x}_{1:t-1} \vert \mathbf{y}_{1:t}\right)\\
    &= p\left(\mathbf{x}_t \vert \mathbf{y}_{1:t}\right) p\left(\mathbf{x}_{1:t-1} \vert \mathbf{y}_{1:t-1}\right)\\
    &= \, \prod_{\tau=1}^t p\left(\mathbf{x}_\tau \vert \mathbf{y}_{1:\tau}\right) \\
    &= \, \prod_{\tau=1}^t \! \mathcal{N}(\mathbf{x}_{\tau}; \mathbf{m}_{{\tau}|1:{\tau}}(\pmb{\theta}), \mathbf{L}_{{\tau}|1:{\tau}}(\pmb{\theta}) ) \\
    & \triangleq p\!\left(\mathbf{x}_{1:t} \vert \mathbf{y}_{1:t}; \pmb{\theta}\right).
\end{array}
\end{eqnarray}
Here we introduced the notation $p(\mathbf{x}_{1:t} | \mathbf{y}_{1:t}; \pmb{\theta} )$ to show the dependency of $p(\mathbf{x}_{1:t} | \mathbf{y}_{1:t} )$ on $\pmb{\theta}$.

We note two interesting issues - a similarity and a difference. Firstly, in \eqref{eq:posterior_update} we notice a similarity with the standard KF update equations for the posterior \cite[Chap. 5]{simon2006optimal}. Secondly, we note a difference. While KF has a Gaussian progression of the posterior to the prior of the next time point, there is no such thing in DANSE. The parameters $\boldsymbol{\alpha}_{t \vert 1:t-1} \left(\pmb{\theta}\right)$ of the prior distribution are obtained from the RNN and the RNN does not use any process model that leads to a Gaussian progression. {A more detailed diagram of the DANSE system is shown in Fig. \ref{fig:danse}(b).}

\subsection{Unsupervised learning of DANSE}\label{sec:unsupervised_learning}
In this subsection, we address how to learn the RNN parameters $\pmb{\theta}$ in DANSE using the training dataset $\mathcal{D} = \lbrace \mathbf{y}^{(i)}_{1:T^{(i)}}\rbrace_{i=1}^N$. We first express 
$p(\mathbf{y}_t | \mathbf{y}_{1:t-1} )$ as follows -
\begin{eqnarray}
\label{eq:pyt_given_prev}
\begin{array}{l}
p(\mathbf{y}_t | \mathbf{y}_{1:t-1} ) 
\\
= \displaystyle\int_{\mathbf{x}_t} p(\mathbf{y}_t | \mathbf{x}_t )p(\mathbf{x}_t |  \mathbf{y}_{1:t-1} ) \, d\mathbf{x}_t \\
 = \!\! \displaystyle\int_{\mathbf{x}_t} \!\! \mathcal{N}(\mathbf{y}_t ;\! \mathbf{H}_t\mathbf{x}_t, \!\mathbf{C}_w \!) \, \mathcal{N}(\mathbf{x}_t; \! \mathbf{m}_{t|1:t-1}(\pmb{\theta}), \! \mathbf{L}_{t|1:t-1}(\pmb{\theta}) \!) \, d\mathbf{x}_t \!\!\!\!\\
 = \mathcal{N}(\mathbf{y}_t ; \mathbf{H}_t \mathbf{m}_{t|1:t-1}(\pmb{\theta}), \mathbf{C}_w + \mathbf{H}_t \mathbf{L}_{t|1:t-1}(\pmb{\theta}) \mathbf{H}_t^{\top}) \\
 \triangleq p(\mathbf{y}_t | \mathbf{y}_{1:t-1}; \pmb{\theta} ).
\end{array}
\end{eqnarray}
In the third line of the above equation, we used \eqref{eq:prior_lik}.
Here we introduced the notation $p(\mathbf{y}_t | \mathbf{y}_{1:t-1}; \pmb{\theta} )$ to show the dependency of $p(\mathbf{y}_t | \mathbf{y}_{1:t-1} )$ on $\pmb{\theta}$. Now, using the chain rule of probability, we can write
\begin{eqnarray}
\label{eq:prob_of_measurements}
\begin{array}{l}
  p\left(\mathbf{y}_{1:T}\right) \triangleq p\left(\mathbf{y}_{1:T}; \pmb{\theta}\right) = \prod_{t=1}^{T}
p\left( \mathbf{y}_t | \mathbf{y}_{1:t-1} ; \pmb{\theta}\right), \, \mathrm{and}  \\
 p\left(\mathbf{y}^{(i)}_{1:T^{(i)}}\right) \triangleq p\left(\mathbf{y}^{(i)}_{1:T^{(i)}}; \pmb{\theta} \right) = \prod_{t=1}^{T^{(i)}}
p\left( \mathbf{y}_t^{(i)} | \mathbf{y}_{1:t-1}^{(i)} ; \pmb{\theta}\!\right)\!\!.
\end{array}
\end{eqnarray}

Therefore, using the dataset $\mathcal{D}$, the maximum-likelihood based unsupervised learning (optmization) problem is

\begin{eqnarray}
\label{eq:optim}
\begin{array}{l}
 \max_{\pmb{\theta}} \log \prod_{i=1}^{N} \prod_{t=1}^{T^{(i)}} p\left( \mathbf{y}_t^{(i)} | \mathbf{y}_{1:t-1}^{(i)} ; {\pmb{\theta}} \right) \\
= \max_{\pmb{\theta}} \sum_{i=1}^{N} \sum_{t=1}^{T^{(i)}} \log p\left( \mathbf{y}_t^{(i)} | \mathbf{y}_{1:t-1}^{(i)} ; {\pmb{\theta}} \right) \\
= \min_{\pmb{\theta}} -\sum_{i=1}^{N} \sum_{t=1}^{T^{(i)}} \log p\left( \mathbf{y}_t^{(i)} | \mathbf{y}_{1:t-1}^{(i)} ; {\pmb{\theta}} \right) \\
= \min_{\pmb{\theta}} \sum_{i=1}^{N} \mathcal{L}\left(\mathbf{y}^{(i)}_{1:T^{(i)}}; {\pmb{\theta}}\right), \\
\end{array}
\end{eqnarray}
with the loss function denoted as $\mathcal{L}$, defined as $\mathcal{L}\left(\mathbf{y}^{(i)}_{1:T^{(i)}}; {\pmb{\theta}}\right) \triangleq - \sum_{t=1}^{T^{(i)}} \log p\left( \mathbf{y}_t^{(i)} | \mathbf{y}_{1:t-1}^{(i)} ; {\pmb{\theta}}\right)$. Dropping the superscipt $(i)$, we can write 
\begin{eqnarray}\label{eq:optim_loss}
\begin{array}{l}
\mathcal{L}\left(\mathbf{y}_{1:T}; {\pmb{\theta}}\right) \triangleq - \sum_{t=1}^{T} \log p\left( \mathbf{y}_t | \mathbf{y}_{1:t-1} ; {\pmb{\theta}}\right) \\
 = \sum_{t=1}^{T} \frac{1}{2} \Vert \mathbf{y}_t - \mathbf{H}_t \mathbf{m}_{t|1:t-1}(\pmb{\theta})\Vert_{\left(\!\mathbf{C}_w + \mathbf{H}_t \mathbf{L}_{t|1:t-1}(\pmb{\theta}) \mathbf{H}_t^{\top}\!\right)^{\!\!-1}}^2\!\!\\
 + \frac{n}{2}\log 2 \pi + \frac{1}{2}\log \text{det} \left(\mathbf{C}_w + \mathbf{H}_t \mathbf{L}_{t|1:t-1}(\pmb{\theta}) \mathbf{H}_t^{\top}\right)\!\!.
\end{array}
\end{eqnarray}
The optimization problem \eqref{eq:optim} can be further formulated as a Bayesian learning problem if we have a suitable prior of ${\pmb{\theta}}$ as $p({\pmb{\theta}})$. For example, if we use an isotropic Gaussian prior, then we have a standard $\ell_2$-norm based regularization in the form of $\| \pmb{\theta} \|_2^2$ penalty. If we use Laplacian prior then we have a sparsity-promoting $\ell_1$-norm based penalty in the form of $\| \pmb{\theta} \|_1$. 
The optimization in \eqref{eq:optim} uses a gradient-descent to learn the parameters $\pmb{\theta}$. 
Therefore, we can comment that the solution approach that we used for the unsupervised learning is a suitable combination of Bayesian learning (or maximum-likelihood learning) and data-driven learning (of RNNs). Fig. \ref{fig:danse}(b) shows a detailed block diagram to depict the training and inference-specific blocks.

{Note that we do not need to use variational inference (VI) to optimize \eqref{eq:optim}. 
VI is an approximate Bayesian learning method, used in dynamical variational autoencoders (DVAEs) \cite{girin2021dynamical, fraccaro2017disentangled, krishnan2017structured}. 
Instead, we endeavor direct optimization of \eqref{eq:optim} using mini-batch gradient-descent, without approximations.} 

\subsubsection*{On the choice of RNNs} We can use prominent RNNs, such as GRUs \cite{choPropertiesNeuralMachine2014} and LSTMs \cite{hochreiter1997long}. Here we used GRUs owing to their simplicity and popularity \cite{chung2014empirical, ravanelli2018ligru}. Specifically, for modeling the mean vector and diagonal covariance matrix of the parameterized Gaussian prior in \eqref{eq:prior_lik}, we transformed the latent state of the GRU using feed-forward networks. We used {softplus activations} to model the non-negative variances. 

\subsection{Bayesian Forecasting {(Solution of T2)}}
We have mentioned the Bayesian forecasting problem in section \ref{subsec:SystemSetup}, as task T2. After unsupervised learning of DANSE, note that \eqref{eq:prior_lik} and \eqref{eq:pyt_given_prev} provide the solution of the forecasting problem T2.
\subsection{Empirical estimation-performance limit of DANSE}
\label{sec:emp_limit}
For a model-free process, it is non-trivial to provide a standard performance limit, for example, quantifying the minimum-mean-square-error (MMSE) performance and comparing it with the practical performance of an estimator. Instead, we can use a pragmatic approach, where we use a supervised learning setup, find a performance limit, and compare. Note that the approach provides an empirical estimation performance limit. The reason for using such a performance limit is an educated guess: an unsupervised learning-based method like DANSE can not perform better than a supervised learning-based method.  

We now explain how to provide the empirical estimation performance limit in the supervised learning setup. Let us assume we have access to a labelled dataset $\overline{\mathcal{D}} \!= \!\left\lbrace \left(\mathbf{x}^{(i)}_{1:T^{(i)}},\mathbf{y}^{(i)}_{1:T^{(i)}}\right)\right\rbrace_{i=1}^N$. Note that $\mathcal{D} \!\subset \! \overline{\mathcal{D}}$. Both $\mathcal{D}$ and $\overline{\mathcal{D}}$ have $N$ training data samples. Dropping the superscript $(i)$, we have state-and-measurement data-pair $\left(\mathbf{x}_{1:T},\mathbf{y}_{1:T}\right)$, that includes $ \left(\mathbf{x}_{1:t},\mathbf{y}_{1:t}\right), t=1,2,\hdots,T$.

There can be many ways to design a supervised learning setup. Let us design a setup where we find parameters of the RNN of DANSE for the Bayesian state estimation $p(\mathbf{x}_{t}|\mathbf{y}_{1:t})$. Using \eqref{eq:posterior_update} and \eqref{eq:posterior_at_time_t}, denoting $p\left( \mathbf{x}_t | \mathbf{y}_{1:t} \right) \triangleq p\left( \mathbf{x}_t | \mathbf{y}_{1:t} ; {{\pmb{\theta}}} \right)$ in \eqref{eq:posterior_at_time_t}, and using the dataset $\overline{\mathcal{D}}$, the maximum-likelihood based supervised learning (optimization) problem is
\begin{eqnarray}
\label{eq:optim_supervised}
\begin{array}{rl}
 \pmb{\theta}_{s} \!\!\!\! & =\arg\max_{\pmb{\theta}} \log \prod_{i=1}^{N} \prod_{t=1}^{T^{(i)}} p\left( \mathbf{x}_t^{(i)} | \mathbf{y}_{1:t}^{(i)} ; {\pmb{\theta}} \right) \\
&= \arg\max_{\pmb{\theta}} \sum_{i=1}^{N} \sum_{t=1}^{T^{(i)}} \log p\left( \mathbf{x}_t^{(i)} | \mathbf{y}_{1:t}^{(i)} ; {\pmb{\theta}} \right) \\
&= \arg\min_{\pmb{\theta}} -\sum_{i=1}^{N} \sum_{t=1}^{T^{(i)}} \log p\left( \mathbf{x}_t^{(i)} | \mathbf{y}_{1:t}^{(i)} ; {\pmb{\theta}} \right) \\
&= \arg\min_{\pmb{\theta}} \sum_{i=1}^{N} \overline{\mathcal{L}}\left( \left( \mathbf{x}^{(i)}_{1:T^{(i)}}, \mathbf{y}^{(i)}_{1:T^{(i)}} \right); {\pmb{\theta}}\right), \\
\end{array}
\end{eqnarray}
with the loss function denoted as $\overline{\mathcal{L}}$, defined as $\overline{\mathcal{L}}\left( \left( \mathbf{x}^{(i)}_{1:T^{(i)}}, \mathbf{y}^{(i)}_{1:T^{(i)}} \right); {\pmb{\theta}}\right) \! \triangleq - \sum_{t=1}^{T^{(i)}} \log p\left( \mathbf{x}_t^{(i)} | \mathbf{y}_{1:t}^{(i)} ; {\pmb{\theta}}\right)$. Dropping the superscript $(i)$, we can write 
\begin{eqnarray} \label{eq:emp_performance_limit_2}
\begin{array}{l}
\overline{\mathcal{L}}\left(\left(\mathbf{x}_{1:T}, \mathbf{y}_{1:T} \right); {\pmb{\theta}}\right) \triangleq - \sum_{t=1}^{T} \log p\left( \mathbf{x}_t | \mathbf{y}_{1:t} ; {\pmb{\theta}}\right) \\
 = \sum_{t=1}^{T} \frac{1}{2} \Vert \mathbf{x}_t - \mathbf{m}_{t|1:t}(\pmb{\theta})\Vert_{\mathbf{L}^{-1}_{t|1:t}(\pmb{\theta})}^2\\
 + \frac{m}{2}\log 2 \pi + \frac{1}{2}\log \text{det} \left( \mathbf{L}_{t|1:t}(\pmb{\theta}) \right).
\end{array}
\end{eqnarray}
Using gradient descent we address the optimization problem \eqref{eq:optim_supervised}, and find $\pmb{\theta}_{s}$; here we use the {subscript} `$s$' to denote `supervised'. A point estimate is $\mathbf{m}_{t|1:t}({\pmb{\theta}_s})$, that provides an empirical MSE performance limit. 


\subsection{On causal and anti-causal usages}
While we developed DANSE maintaining causality, it is amenable for designing anti-causal methods. For example, at time point $t$, we can use inputs $\mathbf{y}_{t+\tau}, \tau \geq 1$ along with $\mathbf{y}_{1:t-1}$ for providing a prior and then find a posterior with the use of $\mathbf{y}_t$. The method can be even fully anti-causal. That means we can use $\mathbf{y}_{t+1:T}$ as an input to provide a prior and then find a posterior with the use of $\mathbf{y}_t$. There can be several possibilities for using measurements in causal and anti-causal manners. In this article, we focus on state estimation maintaining causality.

\section{Experiments and Results}\label{sec:experiments_and_results}
We now perform simulation studies and compare DANSE vis-\`a-vis a few other Bayesian methods, such as model-based KF, EKF, UKF, {a data-driven DMM} and the hybrid method KalmanNet. We use unsupervised KalmanNet \cite{revach2022unsupervised} to have a fair comparison with DANSE. The {six} methods that we compare experimentally have the following properties.
\begin{enumerate}
    \item KF - Knows the linear process model. Optimum for a linear process model. {Knows the measurement system \eqref{eq:measurement}.}
    \item EKF - Knows the linearized form of the nonlinear process model based on a first-order Taylor series expansion. {Knows the measurement system.}
    \item UKF - Knows the nonlinear process model, and uses numerical techniques such as sigma-point methods. {Knows the measurement system.}
    \item Unsupervised KalmanNet - Knows the process model and uses a linearized form of a nonlinear process model, like EKF, but does not need to know the process noise. It knows the measurement system, but does not use the measurement noise statistics. It uses unsupervised learning and the unlabelled training dataset $\mathcal{D}$ {as defined in \eqref{eq:dataset_Def}}.
    \item {DMM-SE - Here we used the DMM method proposed in \cite{krishnan2017structured} to suit causal Bayesian state estimation with the known measurement system \eqref{eq:measurement}. We refer to this method as DMM for Bayesian state estimation (DMM-SE). It has a high connection with the DMM (ST-L) that operates with a `forward' (filtering) inference model structure / utilizing `left' information. It uses unsupervised learning and the unlabelled training dataset $\mathcal{D}$. It does not know the process model but assumes that it is Markovian. It knows the measurement system. The derivation of the optimization problem of DMM-SE is not shown for brevity.}
    \item DANSE - {Does not know the process model. Knows the measurement system and uses the unlabelled training dataset $\mathcal{D}$.} 
\end{enumerate}

For the experiments, our setup is inspired by the KalmanNet papers \cite{revach2022kalmannet, revach2022unsupervised}. We experiment with measurement and state sequences generated from two different SSMs with additive white Gaussian noise. 
We first start with a linear SSM to show a proof-of-concept experiment on how DANSE performs in comparison with the optimal system - the KF. Then we investigate two non-linear SSMs - a Lorenz attractor as used in \cite{revach2022unsupervised} and a Chen attractor \cite{chen1999yet, vcelikovsky2005generalized}. 
For an experiment, we have a training dataset, mentioned before, as $\mathcal{D} = \lbrace \mathbf{y}^{(i)}_{1:T^{(i)}}\rbrace_{i=1}^N$, and a testing dataset $\mathcal{D}_{\text{test}} = \left\lbrace \left(\mathbf{x}^{(j)}_{1:T_{\text{test}}^{(j)}},\mathbf{y}^{(j)}_{1:T_{\text{test}}^{(j)}}\right)\right\rbrace_{j=1}^{\!N_{\text{test}}}$. 
Denoting the estimated state as $\hat{\mathbf{x}}_t$, we use the averaged normalized-mean-squared-error (NMSE) as the performance measure, defined below. 
\begin{equation}
    \text{NMSE} = \frac{1}{N_{\text{test}}} \sum_{j=1}^{N_{\text{test}}} 10\log_{10}\frac{\sum_{t=1}^{T^{(j)}}\Vert\mbx_t^{(j)} - \hat\mbx_t^{(j)}\Vert_2^2}{\sum_{t=1}^{T^{(j)}}\Vert\mbx_t^{(j)}\Vert_2^2}.
\end{equation}
The estimated state $\hat{\mathbf{x}}_t$ corresponds to the posterior mean for all the five state estimation methods considered in this experimental study. 
\subsection{Training, performance measures, and RNN architecture}
We find the architectures for the RNN and the feed-forward networks used in RNN experimentally by grid search and cross-validation. For the RNN, we use a one-layer GRU model with 30 hidden nodes. The feed-forward networks are modeled as shallow networks with a single hidden layer having $32$ hidden nodes. We implement DANSE in Python and PyTorch \cite{paszke2019pytorch} and train the architecture using GPU support. 
The training algorithm uses a mini-batch gradient descent with a batch size of $64$. The optimizer chosen is Adam \cite{kingma2014adam} with an adaptive learning rate set at a starting value $10^{-2}$ and is decreased step-wise by a factor of $0.9$ every ${1/6}'\text{th}$ of the maximum number of training epochs. The maximum number of training epochs is set at $2000$, and early stopping is used to avoid overfitting. In the case of optimizing the empirical estimation performance limit using \eqref{eq:optim_supervised}, we again use Adam as the optimizer with an adaptive learning rate starting at $5 \times 10^{-3}$. The code is available at \begin{small}
\texttt{https://github.com/anubhabghosh/danse\_jrnl} 
\end{small} and \begin{small}
\texttt{https://github.com/saikatchatt/danse-jrnl}.
\end{small}
\subsection{On implementation of the competing methods}
The performances of DANSE are compared with baseline methods that estimate the state sequence from the measurement data.
First, we used a naive, least squares (LS) estimator, which gives estimated state $\hat{\mathbf{x}}_t$ in closed-form as 
\begin{equation}\label{eq:LS}
\hat{\mathbf{x}}_t = \left(\mathbf{H}_t^{\top} \mathbf{C}_w^{-1} \mathbf{H}_t\right)^{-1} \mathbf{H}_t^{\top} \mathbf{C}_w^{-1} \mathbf{y}_t
\end{equation}
We implemented KF, EKF, and UKF using PyTorch and FilterPy \cite{labbe2020filterpy}. 
We used the implementation of the unsupervised KalmanNet from \cite{revach2022unsupervised} for our experimental study. {DMM-SE was implemented following the derivations in \cite{krishnan2017structured}, \cite[Chap. 5]{girin2021dynamical}. The code for DMM was adapted in PyTorch based on the original implementation in \cite{krishnan2017structured}.}


\subsection{Linear SSM}\label{sec:linear_ssm}
To compare DANSE with the optimal KF, we first perform a proof-of-concept experiment with a linear SSM. The experiment is motivated by \cite{revach2022unsupervised, revach2022kalmannet}. The linear SSM model is as follows: 
\begin{equation}\label{eq:linearstatespace}
\begin{aligned}
    \mbx_{t}&=\mathbf{F}\mbx_{t-1} + \mathbf{e}_t \in \mathbb{R}^2, \\
    \mby_{t}&=\mathbf{H}\mbx_{t} + \mathbf{w}_t \in \mathbb{R}^2,
\end{aligned}
\end{equation}
where $t=1,2,\dots,T$, $\mathbf{F}\in\R^{2\times 2}$, $\mathbf{H}\in\R^{2\times 2}$, $\mathbf{e}_t\sim\mathcal{N}(\boldsymbol{0},\mathbf{C}_e)$ is the white Gaussian process noise variable with $\mathbf{C}_e=\sigma_e^2 \mathbf{I}_{2}$, and $\mathbf{w}_t\sim\mathcal{N}(\boldsymbol{0},\mathbf{C}_w)$ is the white Gaussian measurement noise with $\mathbf{C}_w=\sigma_w^2 \mathbf{I}_{2}$ ($\mathbf{I}_2$ denotes the $2 \times 2$ identity matrix).
The matrices $\mathbf{F}, \mathbf{H}$ are time-invariant, where $\mathbf{H}$ is an identity matrix and $\mathbf{F}$ is a scaled upper triangular matrix.   

We simulate \eqref{eq:linearstatespace} to create the training and testing data, and then compare DANSE with the standard KF and LS. LS does not use time-wise correlation. KF knows the signal model. All of them know the measurement system and measurement noise statistics. 
We train DANSE using a training dataset $\mathcal{D}$, where $N=500, T=500$ (all the training sequences are of the same length). 
We {vary} the signal-to-measurement noise ratio (SMNR) from $-10$ to $30$ dB, with a fixed $\sigma_e^2=-10 \text{ dB}$. 
The SMNR is calculated on $\mathcal{D}_{\text{test}}$ as
\begin{eqnarray}\label{eq:SMNR}
\begin{array}{lr}
    \text{SMNR} \!\!\!\!\!\!&= \!\! \frac{1}{N_{\text{test}}} \!\! \sum \limits_{j=1}^{N_{\text{test}}}  \!\! 10 \log_{10} \!\!\left( \frac{\sum_{t=1}^{T_{\text{test}}^{(j)}} \mathbb{E}\lbrace  \Vert \mathbf{H}\mbx_t^{(j)} - \mathbb{E}\lbrace \mathbf{H} \mbx_t^{(j)} \rbrace \Vert_2^2 \rbrace}{\text{tr}\left(\mathbf{C}_w\right)}\right) \\
    \!\!&= \!\! \frac{1}{N_{\text{test}}} \!\! \sum\limits_{j=1}^{N_{\text{test}}} \!\! 10 \log_{10} \!\! \left( \frac{\sum_{t=1}^{T_{\text{test}}^{(j)}} \mathbb{E}\lbrace  \Vert \mathbf{H}\mbx_t^{(j)} - \mathbb{E}\lbrace \mathbf{H} \mbx_t^{(j)} \rbrace \Vert_2^2 \rbrace}{n\sigma_w^2}\right) \\
\end{array}
\end{eqnarray}

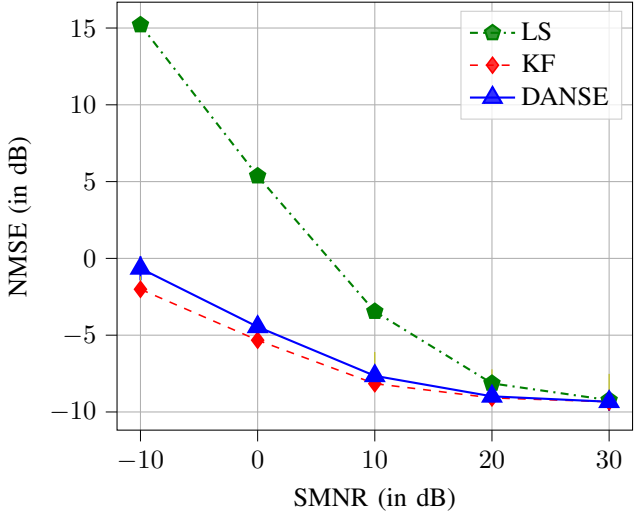
\begin{figure}[t]
    \centering
    \scalebox{1.0}{
\begin{tikzpicture}

\definecolor{darkgray176}{RGB}{176,176,176}
\definecolor{goldenrod1911910}{RGB}{191,191,0}
\definecolor{green01270}{RGB}{0,127,0}
\definecolor{lightgray204}{RGB}{204,204,204}

\begin{axis}[
legend cell align={left},
legend style={fill opacity=0.8, draw opacity=1, text opacity=1, draw=lightgray204},
tick align=outside,
tick pos=left,
x grid style={darkgray176},
xlabel={SMNR (in dB)},
xmajorgrids,
xmin=-12, xmax=32,
xtick style={color=black},
y grid style={darkgray176},
ylabel={NMSE (in dB)},
ymajorgrids,
ymin=-27.2955385431647, ymax=17.4491397723556,
ytick style={color=black}
]
\path [draw=green01270, semithick]
(axis cs:-10,14.9891453534365)
--(axis cs:-10,15.4152907580137);

\path [draw=green01270, semithick]
(axis cs:0,5.01354393362999)
--(axis cs:0,5.39426437020302);

\path [draw=green01270, semithick]
(axis cs:10,-4.99082292616367)
--(axis cs:10,-4.57218824326992);

\path [draw=green01270, semithick]
(axis cs:20,-14.9955393373966)
--(axis cs:20,-14.5621846616268);

\path [draw=green01270, semithick]
(axis cs:30,-24.9790792912245)
--(axis cs:30,-24.6092881709337);

\path [draw=red, semithick]
(axis cs:-10,-2.88632434606552)
--(axis cs:-10,-2.21148735284805);

\path [draw=red, semithick]
(axis cs:0,-7.1343582868576)
--(axis cs:0,-6.5706502199173);

\path [draw=red, semithick]
(axis cs:10,-11.9392495453358)
--(axis cs:10,-11.3219732940197);

\path [draw=red, semithick]
(axis cs:20,-17.0818084031343)
--(axis cs:20,-16.5830361098051);

\path [draw=red, semithick]
(axis cs:30,-25.2616895288229)
--(axis cs:30,-24.8897005468607);

\path [draw=blue, thick]
(axis cs:-10,4.93231421709061)
--(axis cs:-10,6.64130848646164);

\path [draw=blue, thick]
(axis cs:0,-3.52420926094055)
--(axis cs:0,-3.00474858283997);

\path [draw=blue, thick]
(axis cs:10,-10.2813050746918)
--(axis cs:10,-9.65643525123596);

\path [draw=blue, thick]
(axis cs:20,-17.0053861886263)
--(axis cs:20,-16.5071298331022);

\path [draw=blue, thick]
(axis cs:30,-25.2607380598783)
--(axis cs:30,-24.8876422196627);

\addplot [thick, green01270, dashdotted, mark=pentagon*, mark size=3, mark options={solid}]
table {%
-10 15.2022180557251
0 5.2039041519165
10 -4.7815055847168
20 -14.7788619995117
30 -24.7941837310791
};
\addlegendentry{LS}
\addplot [semithick, red, dashed, mark=diamond*, mark size=3, mark options={solid}]
table {%
-10 -2.54890584945679
0 -6.85250425338745
10 -11.6306114196777
20 -16.8324222564697
30 -25.0756950378418
};
\addlegendentry{KF}
\addplot [thick, blue, mark=triangle*, mark size=4, mark options={solid}]
table {%
-10 5.78681135177612
0 -3.26447892189026
10 -9.96887016296387
20 -16.7562580108643
30 -25.0741901397705
};
\addlegendentry{DANSE}
\end{axis}

\end{tikzpicture}}
    \caption{Plot of NMSE (in dB) vs SMNR (in dB) for the linear SSM. Here, KF performance is the lower bound and LS is a possible upper bound.} 
    \label{fig:nmse_linear_ssm}
\end{figure}

We evaluate all the methods on the same test set. For the test set, we have $N_{\text{test}}=100$ trajectories with $T_{\text{test}}=1000$. We deliberately use the longer test-sample trajectory than the training-sample trajectory to check whether DANSE can generalize to longer trajectory lengths. The variation of NMSE versus SMNR is shown in Fig. \ref{fig:nmse_linear_ssm}. We can observe that DANSE performs close to the optimal KF. Note that the gap between KF and DANSE consistently increases as SMNR decreases. This is a consistent result for learning with noisy data. 
Moreover, we observed empirically that the gap between KF and DANSE decreases with the use of more training data samples $N$ of $\mathcal{D}$.
This is consistent with the behavior we expected, and we do not elaborate further on the sample complexity for the linear case.

\begin{figure*}[t]
\centering
  \includegraphics[width=\textwidth]{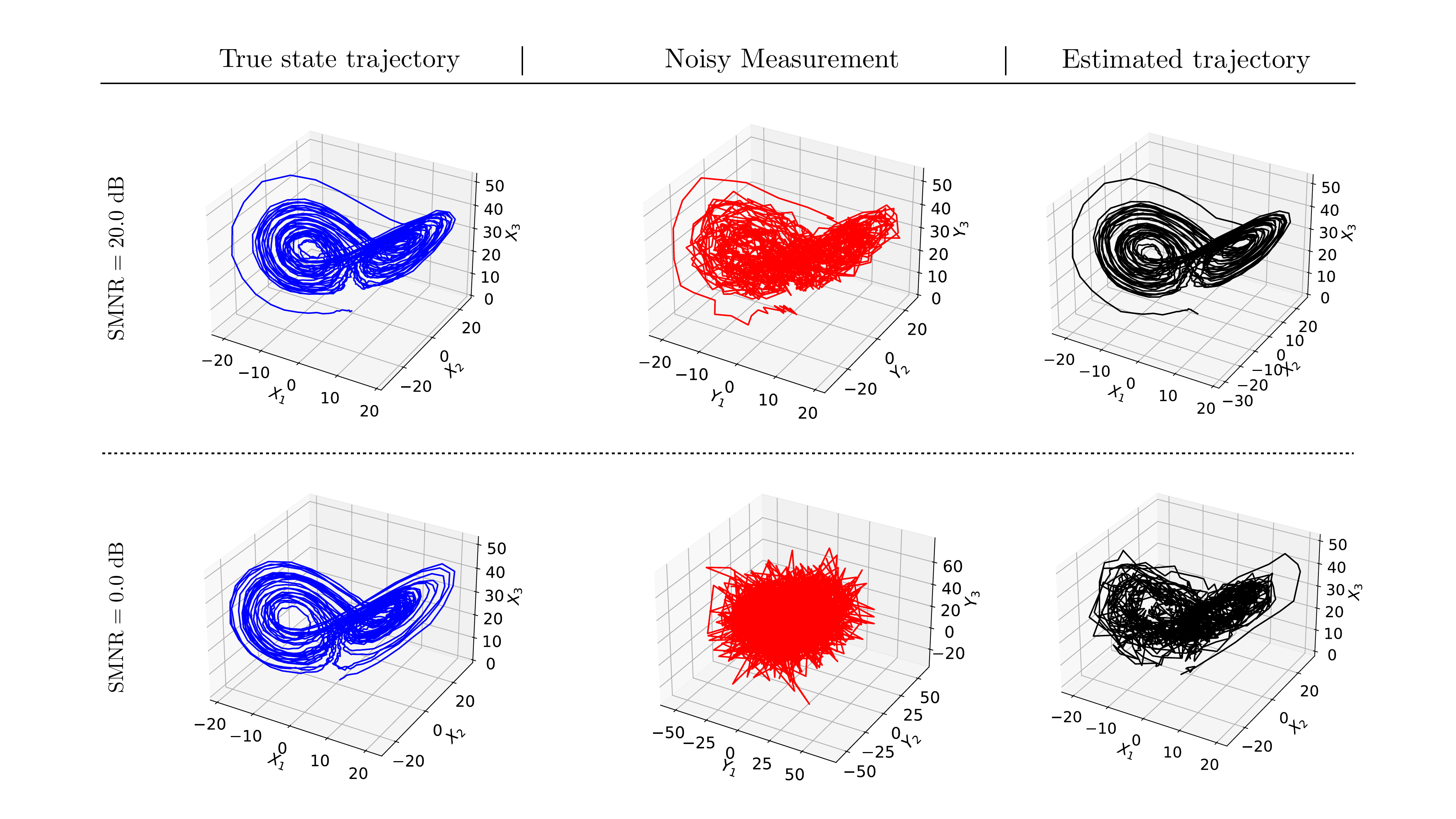}
  \caption{{State estimation for Lorenz attractor SSM using DANSE, at 20 dB and 0 dB SMNR, with $\mathbf{H}_t = \mathbf{I}_3$. The first row is for 20 dB SMNR, and the second row is for 0 dB SMNR. Note that, in unsupervised learning, at a chosen SMNR, a training dataset $\mathcal{D}$ comprises of $N$ such `red colored' noisy measurement trajectories as data samples.}}
  \label{fig:lorenzssm_demonstrator}
\end{figure*}


\begin{figure}[t]
\centering
\scalebox{1.0}{
\begin{tikzpicture}

\definecolor{darkgray176}{RGB}{176,176,176}
\definecolor{darkturquoise0191191}{RGB}{0,191,191}
\definecolor{darkviolet1910191}{RGB}{191,0,191}
\definecolor{goldenrod1911910}{RGB}{191,191,0}
\definecolor{green01270}{RGB}{0,127,0}
\definecolor{lightgray204}{RGB}{204,204,204}

\begin{axis}[
legend cell align={left},
legend style={fill opacity=0.8, draw opacity=1, text opacity=1, draw=lightgray204},
tick align=outside,
tick pos=left,
x grid style={darkgray176},
xlabel={SMNR (in dB)},
xmajorgrids,
xmin=-12, xmax=32,
xtick style={color=black},
y grid style={darkgray176},
ylabel={NMSE (in dB)},
ymajorgrids,
ymin=-36.8098631691188, ymax=1.1404801573604,
ytick style={color=black}
]

\path [draw=red, semithick]
(axis cs:-10,-2.79836767911911)
--(axis cs:-10,-1.6387203335762);

\path [draw=red, semithick]
(axis cs:0,-10.1341648101807)
--(axis cs:0,-7.60629272460938);

\path [draw=red, semithick]
(axis cs:10,-22.9056630432606)
--(axis cs:10,-22.2137102782726);

\path [draw=red, semithick]
(axis cs:20,-28.9034972637892)
--(axis cs:20,-28.5270142108202);

\path [draw=red, semithick]
(axis cs:30,-34.6303021088243)
--(axis cs:30,-34.4018176421523);

\path [draw=black, semithick]
(axis cs:-10,-6.70856255292892)
--(axis cs:-10,-6.13578635454178);

\path [draw=black, semithick]
(axis cs:0,-14.2738043069839)
--(axis cs:0,-11.8986238241196);

\path [draw=black, semithick]
(axis cs:10,-22.8868536651134)
--(axis cs:10,-22.2221818268299);

\path [draw=black, semithick]
(axis cs:20,-28.8260292410851)
--(axis cs:20,-28.4580970406532);

\path [draw=black, semithick]
(axis cs:30,-34.4913239628077)
--(axis cs:30,-34.2688841670752);

\path [draw=blue, thick]
(axis cs:-10,-5.19774712622166)
--(axis cs:-10,-4.81431113183498);

\path [draw=blue, thick]
(axis cs:0,-13.1789858937263)
--(axis cs:0,-12.1048256754875);

\path [draw=blue, thick]
(axis cs:10,-20.8483796715736)
--(axis cs:10,-20.2195066809654);

\path [draw=blue, thick]
(axis cs:20,-27.7324056625366)
--(axis cs:20,-27.2645235061646);

\path [draw=blue, thick]
(axis cs:30,-34.2813821136951)
--(axis cs:30,-33.9816565215588);

\path [draw=goldenrod1911910, thick]
(axis cs:-10,-5.5986662954092)
--(axis cs:-10,-5.36187066137791);

\path [draw=goldenrod1911910, thick]
(axis cs:0,-6.20716931670904)
--(axis cs:0,-6.01917574554682);

\path [draw=goldenrod1911910, thick]
(axis cs:10,-12.4230005592108)
--(axis cs:10,-12.0925176292658);

\path [draw=goldenrod1911910, thick]
(axis cs:20,-21.3682191669941)
--(axis cs:20,-21.0524923503399);

\path [draw=goldenrod1911910, thick]
(axis cs:30,-26.9404115900397)
--(axis cs:30,-26.7176656499505);

\path [draw=darkturquoise0191191, thick]
(axis cs:-10,-6.90227784216404)
--(axis cs:-10,-6.51884184777737);

\path [draw=darkturquoise0191191, thick]
(axis cs:0,-14.0925391316414)
--(axis cs:0,-13.0183789134026);

\path [draw=darkturquoise0191191, thick]
(axis cs:10,-22.2210698723793)
--(axis cs:10,-21.5921968817711);

\path [draw=darkturquoise0191191, thick]
(axis cs:20,-28.4851655960083)
--(axis cs:20,-28.0172834396362);

\path [draw=darkturquoise0191191, thick]
(axis cs:30,-34.441809207201)
--(axis cs:30,-34.1420836150646);

\path [draw=goldenrod1911910]
(axis cs:-10,-3.40908762812614)
--(axis cs:-10,-2.77661564946175);

\path [draw=goldenrod1911910]
(axis cs:0,-11.5875419378281)
--(axis cs:0,-10.2184063196182);

\path [draw=goldenrod1911910]
(axis cs:10,-19.5977766215801)
--(axis cs:10,-19.1177842915058);

\path [draw=goldenrod1911910]
(axis cs:20,-27.3844678252935)
--(axis cs:20,-27.0147745758295);

\path [draw=goldenrod1911910]
(axis cs:30,-33.6548755988479)
--(axis cs:30,-33.4152232781053);

\addplot [semithick, red, dashed, mark=diamond*, mark size=3, mark options={solid}]
table {%
-10 -2.21854400634766
0 -8.87022876739502
10 -22.5596866607666
20 -28.7152557373047
30 -34.5160598754883
};
\addlegendentry{EKF}
\addplot [semithick, black, mark=*, mark size=3, mark options={solid}]
table {%
-10 -6.42217445373535
0 -13.0862140655518
10 -22.5545177459717
20 -28.6420631408691
30 -34.3801040649414
};
\addlegendentry{UKF}
\addplot [thick, darkviolet1910191, mark=square, mark size=3, mark options={solid}]
table {%
-10 -5.48026847839355
0 -6.11317253112793
10 -12.2577590942383
20 -21.210355758667
30 -26.8290386199951
};
\addlegendentry{DMM - SE}
\addplot [goldenrod1911910, mark=square*, mark size=3, mark options={solid}]
table {%
-10 -3.09285163879395
0 -10.9029741287231
10 -19.357780456543
20 -27.1996212005615
30 -33.5350494384766
};
\addlegendentry{KalmanNet}
\addplot [thick, dashed, darkturquoise0191191, mark=triangle, mark size=4, line width=1.1, mark options={solid}]
table {%
-10 -6.7105598449707
0 -13.555459022522
10 -21.9066333770752
20 -28.2512245178223
30 -34.2919464111328
};
\addlegendentry{Emp. limit}
\addplot [thick, blue, mark=triangle*, mark size=3, line width=1.1, mark options={solid}]
table {%
-10 -5.00602912902832
0 -12.6419057846069
10 -20.5339431762695
20 -27.4984645843506
30 -34.131519317627
};
\addlegendentry{DANSE}
\end{axis}

\end{tikzpicture}}
\caption{{Performance comparison of all methods for the Lorenz attractor SSM, including the empirical estimation performance limit. DANSE, KalmanNet, {DMM-SE} and the empirical estimation performance limit {(denoted in the figure by `Emp. limit')} were all trained using $\mathcal{D}$ with $N=1000, T=100$.}}
\label{fig:nmse_lorenz_full_limit}
\end{figure}
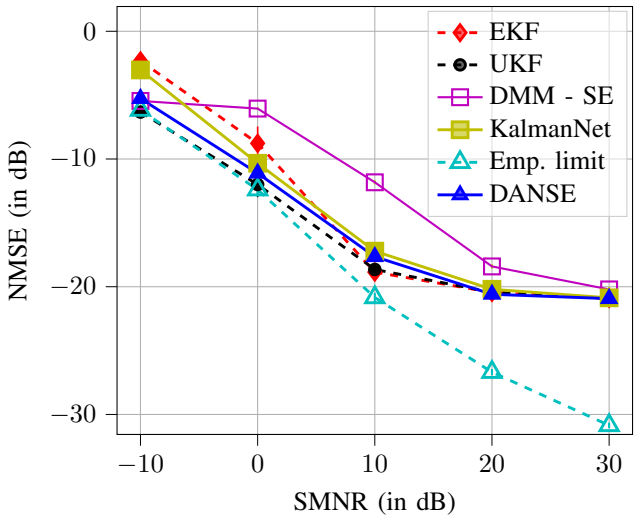
\subsection{Lorenz attractor SSM (nonlinear SSM)} \label{sec:Lorenz_SSM}
Next, we experiment with a non-linear SSM - the Lorenz attractor SSM \cite{lorenz1963deterministic}. Similarly to \cite{revach2022kalmannet, garcia2019combining}, the SSM is in a discretized form as 
\begin{equation}\label{eq:lorenz_state}
    \mbx_{t}=\mathbf{F}_t(\mbx_{t-1})\mbx_{t-1} + \mathbf{e}_t \in \mathbb{R}^3,
\end{equation}
where $\mathbf{e}_t\sim\mathcal{N}(\boldsymbol{0},\mathbf{C}_e)$ with $\mathbf{C}_e=\sigma_e^2 \mathbf{I}_{3}$, and
\begin{eqnarray}\label{eq:F_lorenz}
\mathbf{F}_t(\mbx_{t-1}) = \exp\left(\begin{bmatrix}
-10 & 10 & 0 \\
28 & -1 & -x_{t-1, 1} \\
0 & x_{t-1, 1} & -\frac{8}{3} \\ 
\end{bmatrix}\Delta \right),
\end{eqnarray}
with the step-size $\Delta=0.02 \text{ seconds}$. In our simulations, we use a finite-Taylor series approximation of $5'\text{th}$ order for \eqref{eq:F_lorenz}. The measurement system is
\begin{equation}
   \mby_{t} = \mathbf{H}_t\mbx_{t} + \mathbf{w}_t \in \mathbb{R}^3. 
\end{equation}
We use $\mathbf{H}_t=\mathbf{I}_3$. The measurement noise is $\mathbf{w}_t\sim\mathcal{N}(\boldsymbol{0},\mathbf{C}_w)$ with $\mathbf{C}_w=\sigma_w^2 \mathbf{I}_{3}$, while $\sigma_e^2$ corresponds to $-10$\text{ dB}. For this SSM, we compare DANSE vis-à-vis EKF, UKF, {DMM-SE} and unsupervised KalmanNet. {We train DANSE, {DMM-SE} and KalmanNet using dataset $\mathcal{D}$ with $N=1000$, $T=100$.} 
We evaluate all the methods on the same test set $\mathcal{D}_{\text{test}}$ with $N_{\text{test}}=100$ and $T_{\text{test}}=2000$. Like the linear SSM case, we test on longer trajectories than the training.

{While we have used $\mathbf{H}_t=\mathbf{I}_3$ for experiments, we could have used any $3 \times 3$ full rank matrix as a measurement matrix, and the performance trend is expected to be same. The reason is that we can pre-multiply both sides of \eqref{eq:measurement} by $\mathbf{H}_t^{-1}$ and transform the  measurement vector as $\mathbf{y}_t'= \mathbf{x}_t + \mathbf{w}_t'$, where $\mathbf{y}_t' \triangleq \mathbf{H}_t^{-1}\mathbf{y}_t$ and $\mathbf{w}_t' \triangleq \mathbf{H}_t^{-1}\mathbf{w}_t \sim \mathcal{N}\left(\boldsymbol{0}, \mathbf{C}_{w'}\right) =  \mathcal{N}\left(\boldsymbol{0}, \mathbf{H}_t^{-1}\mathbf{C}_{w} (\mathbf{H}_t^{-1})^{\top}\right)$. We could provide the same argument for a full column rank matrix $\mathbf{H}_t$. In that case we will use a left pseudo-inverse $\mathbf{H}_t^{\dagger} \triangleq \left(\mathbf{H}_t^{\top}\mathbf{H}_t\right)^{-1}\mathbf{H}_t^{\top}$ instead of $\mathbf{H}_t^{-1}$.}

\subsubsection{State estimation}
With a focus on the state estimation task, we start with a visual illustration of the DANSE performance. Fig. \ref{fig:lorenzssm_demonstrator} shows two random instances of Lorenz attractor SSM trajectories, their noisy measurements, and corresponding estimates using DANSE. The top row is SMNR = 20 dB and the bottom row for SMNR = 0 dB. Note that, at a particular SMNR, the training dataset $\mathcal{D}$ is comprised of such kind of noisy measurement trajectories. Even at SMNR = 0 dB, without knowing the process model, DANSE is able to extract meaningful information from the noisy input and then provide a reasonable estimate of the state trajectory. 

\begin{figure}[t]
    \centering
    \scalebox{1.05}{\input{figs/Trajectories_sigma_e2_-10.0dB_smnr_0.0dB_Ver2}}
    \caption{Visualization of mean and uncertainty of the estimated posterior $x_{t,2}$ given $\mathbf{y}_{1:t}$, that means $p(x_{t,2}|\mathbf{y}_{1:t})$, for the Lorenz attractor SSM. We have $\sigma_e^2$ corresponding to $-10$ dB, and $\text{SMNR}=0$ dB. The shaded areas in red and yellow show the uncertainty of the estimate by one-sigma-point using DANSE and UKF, respectively. {DANSE is trained using a dataset $\mathcal{D}$ with $N=1000, T=100$}. 
    }
    \label{fig:trajectories_lorenz_ssm}
\end{figure}
Then we show a performance comparison (NMSE versus SMNR) for all the competing methods in Fig. \ref{fig:nmse_lorenz_full_limit}. In the figure, we also show the empirical estimation performance limit of DANSE, mentioned in subsection \ref{sec:emp_limit}. We can see that, across different SMNR values, DANSE provides competitive performances. {The performance of DMM-SE is poorer compared to the other methods in the low SMNR region and this is in agreement with the findings in \cite[Appendix D]{krishnan2017structured}, where they show that causal variations of DMM such as DMM (ST-L) underperforms on a nonlinear state estimation task compared to model-based methods such as UKF.} 



We now visually show the uncertainty of estimates using a single-dimensional test set trajectory. In Fig. \ref{fig:nmse_lorenz_full_limit}, at 0 dB SMNR, UKF is found to be the most competitive to DANSE in terms of NMSE. This motivates us to qualitatively compare the state estimates of UKF and DANSE. In Fig. \ref{fig:trajectories_lorenz_ssm}, we show the visualization of the estimated posterior of $x_2$-dimension of three-dimensional Lorenz attractor SSM, {using a single test set trajectory}. The shaded regions represent the uncertainties of the estimates by one-sigma-point (standard deviation of the posterior). It is clear that DANSE has less uncertainty than the UKF.

\subsubsection{{Forecasting}} 
\begin{figure}[t]
    \centering
    \scalebox{1.05}{\input{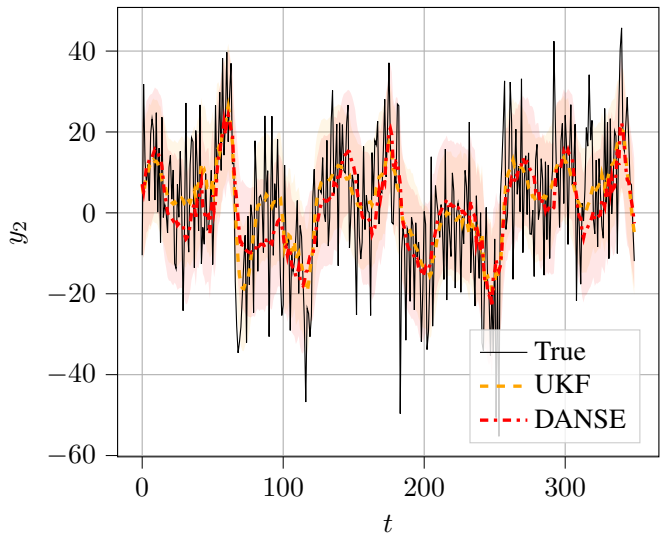}}
    \caption{Visualization of the mean and uncertainty of the forecast of ${y}_{t,2}$ given $\mathbf{y}_{1:t-1}$, that means $p({y}_{t,2} | \mathbf{y}_{1:t-1})$, for the  Lorenz attractor SSM. We have $\sigma_e^2$ corresponding to $-10$ dB, and $\text{SMNR} = 0$ dB. The shaded areas in red and yellow show the uncertainty of the forecast by one-sigma-point using DANSE and UKF respectively. {DANSE is trained using a dataset $\mathcal{D}$ with $N=1000, T=100$}.
    }
    \label{fig:meas_trajectories_lorenz_ssm}
\end{figure}

In this subsection, we show forecasting results visually for the Lorenz attractor SSM. We show uncertainty of forecasting 
$p(\mathbf{y}_{t+1}|\mathbf{y}_{1:t})$ across single-dimensional trajectories. We use the same experimental conditions as Fig. \ref{fig:trajectories_lorenz_ssm} in the previous experiment, and show the second-dimension of the three-dimensional trajectories, i.e. $p\left(y_{t,2}|\mathbf{y}_{1:t-1}\right)$. 
UKF and DANSE are compared in relation to the true trajectories. Fig. \ref{fig:meas_trajectories_lorenz_ssm} shows the forecasting of ${y}_{t,2}$, i.e. $p({y}_{t,2}|\mathbf{y}_{1:t-1})$ at 0 dB SMNR. 
The shaded regions represent the uncertainty of the estimates by one-sigma-point. Note that $\mathbf{y}_t$ is difficult to forecast at the low SMNR = 0 dB, as $\mathbf{y}_t$ has a significant amount of noise component.

\subsubsection{Remark} Using the above experiments and visualizations, we demonstrate that the DANSE is competitive. The results show that it is possible to design an unsupervised learning-based data-driven method to estimate the non-linear state of a process-free model using linear measurements.

\subsection{On training data amount and mismatched conditions}
Given the encouraging performance of DANSE, two important questions arise: (a) How much training data is required for DANSE to perform well? We do not have any theoretical results for this question. Instead, we deliberate on the question using experiments. (b) All the experiments so far mentioned are conducted in matched training-and-testing conditions. What will happen in mismatched conditions? This question is central to a standard robustness study. We deliberate on these two questions in the following. {We choose KalmanNet for comparison, which is training-based and seems to yield performance close to DANSE as per Fig. \ref{fig:nmse_lorenz_full_limit}.}
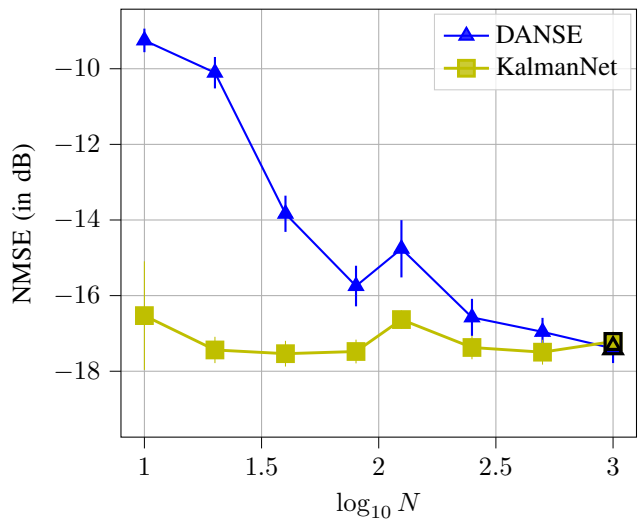
\begin{figure}[t]
\centering
    \scalebox{1.0}{
\begin{tikzpicture}

\definecolor{darkgray176}{RGB}{176,176,176}
\definecolor{goldenrod1911910}{RGB}{191,191,0}
\definecolor{green01270}{RGB}{0,127,0}
\definecolor{lightgray204}{RGB}{204,204,204}

\begin{axis}[
legend cell align={left},
legend style={
  fill opacity=0.8,
  draw opacity=1,
  text opacity=1,
  at={(0.99,0.9)},
  anchor=east,
  draw=lightgray204
},
tick align=outside,
tick pos=left,
x grid style={darkgray176},
xlabel={\(\displaystyle \log_{10} N\)},
xmajorgrids,
xmin=0.9, xmax=3.1,
xtick style={color=black},
y grid style={darkgray176},
ylabel={NMSE (in dB)},
ymajorgrids,
ymin=-19.7428541466594, ymax=-8.42484700828791,
ytick style={color=black}
]

\path [draw=blue, thick]
(axis cs:1,-9.56190738081932)
--(axis cs:1,-8.93930187821388);

\path [draw=blue, thick]
(axis cs:1.30102999566398,-10.5200670063496)
--(axis cs:1.30102999566398,-9.68888494372368);

\path [draw=blue, thick]
(axis cs:1.60205999132796,-14.3145996630192)
--(axis cs:1.60205999132796,-13.3565448224545);

\path [draw=blue, thick]
(axis cs:1.90308998699194,-16.283363878727)
--(axis cs:1.90308998699194,-15.210029065609);

\path [draw=blue, thick]
(axis cs:2.09691001300806,-15.5186216235161)
--(axis cs:2.09691001300806,-14.0039346814156);

\path [draw=blue, thick]
(axis cs:2.39794000867204,-17.0654890239239)
--(axis cs:2.39794000867204,-16.0856569111347);

\path [draw=blue, thick]
(axis cs:2.69897000433602,-17.3279871344566)
--(axis cs:2.69897000433602,-16.5885587334633);

\path [draw=blue, thick]
(axis cs:3,-17.7836344540119)
--(axis cs:3,-17.011329382658);

\path [draw=goldenrod1911910]
(axis cs:1,-17.9644935131073)
--(axis cs:1,-15.09308552742);

\path [draw=goldenrod1911910]
(axis cs:1.30102999566398,-17.7848579585552)
--(axis cs:1.30102999566398,-17.0895431339741);

\path [draw=goldenrod1911910]
(axis cs:1.60205999132796,-17.8751656115055)
--(axis cs:1.60205999132796,-17.1960280835629);

\path [draw=goldenrod1911910]
(axis cs:1.90308998699194,-17.792289942503)
--(axis cs:1.90308998699194,-17.1641538441181);

\path [draw=goldenrod1911910]
(axis cs:2.09691001300806,-16.8746770769358)
--(axis cs:2.09691001300806,-16.3928018659353);

\path [draw=goldenrod1911910]
(axis cs:2.39794000867204,-17.6830503940582)
--(axis cs:2.39794000867204,-17.060907125473);

\path [draw=goldenrod1911910]
(axis cs:2.69897000433602,-17.8248949050903)
--(axis cs:2.69897000433602,-17.1666173934937);

\path [draw=goldenrod1911910]
(axis cs:3,-17.4753162562847)
--(axis cs:3,-16.9522655308247);

\addplot [thick, blue, mark=triangle*, mark size=3, mark options={solid}]
table {%
1 -9.2506046295166
1.30102999566398 -10.1044759750366
1.60205999132796 -13.8355722427368
1.90308998699194 -15.746696472168
2.09691001300806 -14.7612781524658
2.39794000867204 -16.5755729675293
2.69897000433602 -16.95827293396
3 -17.397481918335
};
\addlegendentry{DANSE}
\addplot [goldenrod1911910, mark=square*, mark size=3, line width=1.1, mark options={solid}]
table {%
1 -16.5287895202637
1.30102999566398 -17.4372005462646
1.60205999132796 -17.5355968475342
1.90308998699194 -17.4782218933105
2.09691001300806 -16.6337394714355
2.39794000867204 -17.3719787597656
2.69897000433602 -17.495756149292
3 -17.2137908935547
};
\addlegendentry{KalmanNet}
\addplot[very thick, mark=triangle, mark size=4, mark options={solid}] coordinates {(3, -17.397481918335)};
\addplot[very thick, mark=square, mark size=3, mark options={solid}] coordinates {(3, -17.2137908935547)};
\end{axis}

\end{tikzpicture}}
\caption{{Effect of increasing the number of samples $N$ on the performance of DANSE and KalmanNet. The black-bordered marker indicates the value of $N=1000$, $T=100$ used throughout the paper for training DANSE and KalmanNet.}} 
\label{fig:1d_lorenz_diff_N}
\end{figure}
\subsubsection{On the training data amount}
We wish to study the effect of the number of sample trajectories required for training DANSE. For this study, we vary the number of sample trajectories - $N$ in $\mathcal{D}$, which means the size of $\mathcal{D}$. We keep the length of a trajectory $T$ fixed at $T=100$ and perform the experiments
We also have a fixed $\sigma_e^2=-10$ dB, and $\text{SMNR}=10$ dB. We train DANSE at varying $N$ and evaluate on a test dataset $\mathcal{D}_{\text{test}}$ with $N_{\text{test}}=100, T_{\text{test}}=2000$ at the same SMNR and $\sigma_e^2$ as for training. The results are shown in Fig. \ref{fig:1d_lorenz_diff_N}. We see that DANSE requires around $N=1000$ training sample trajectories to provide good performance. The performance improves with an approximately linear trend with respect to $\log_{10} N$ for small $N$ and then shows a saturation trend. Overall, the performance indicates an (approximate) exponential trend with respect to $N$ -- a power law behavior. This reminds us of usual trends in performance-versus-resource curves, such as distortion-rate curves in source coding. Note that we include KalmanNet's behavior in the same figure and find that it can be trained with fewer data. We believe that the reason is -- KalmanNet knows the process SSM. On the other hand, DANSE requires more data as it does not know the process SSM.

\subsubsection{On mismatched conditions for robustness study}\label{sec:mismatched_robustness_study}
\begin{figure}[t]
    \centering
    \scalebox{1.0}{
\begin{tikzpicture}

\definecolor{darkgray176}{RGB}{176,176,176}
\definecolor{goldenrod1911910}{RGB}{191,191,0}
\definecolor{green01270}{RGB}{0,127,0}
\definecolor{lightgray204}{RGB}{204,204,204}

\begin{axis}[
legend cell align={left},
legend style={
  fill opacity=0.8,
  draw opacity=1,
  text opacity=1,
  at={(0.01,0.9)},
  anchor=west,
  draw=lightgray204
},
tick align=outside,
tick pos=left,
x grid style={darkgray176},
xlabel={\(\displaystyle \sigma_e^2\) (in dB)},
xmajorgrids,
xmin=-21.25, xmax=6.25,
xtick style={color=black},
y grid style={darkgray176},
ylabel={NMSE (in dB)},
ymajorgrids,
ymin=-22.0292878992856, ymax=-10.462478300184,
ytick style={color=black}
]

\path [draw=blue, thick]
(axis cs:-20,-21.5035238265991)
--(axis cs:-20,-20.8821039199829);

\path [draw=blue, thick]
(axis cs:-10,-20.8465705215931)
--(axis cs:-10,-20.2163223922253);

\path [draw=blue, thick]
(axis cs:-5,-19.5686340928078)
--(axis cs:-5,-18.9637221693993);

\path [draw=blue, thick]
(axis cs:0,-17.1191830039024)
--(axis cs:0,-16.5482440590858);

\path [draw=blue, thick]
(axis cs:5,-13.6916997432709)
--(axis cs:5,-13.1109125614166);

\path [draw=goldenrod1911910]
(axis cs:-20,-19.7916278541088)
--(axis cs:-20,-19.3521900475025);

\path [draw=goldenrod1911910]
(axis cs:-10,-19.5956939160824)
--(axis cs:-10,-19.1173149645329);

\path [draw=goldenrod1911910]
(axis cs:-5,-19.193228662014)
--(axis cs:-5,-18.6284494996071);

\path [draw=goldenrod1911910]
(axis cs:0,-18.0392674505711)
--(axis cs:0,-17.5042963922024);

\path [draw=goldenrod1911910]
(axis cs:5,-15.4084194302559)
--(axis cs:5,-14.5454456210136);

\addplot [thick, cyan, dashdotted, mark=triangle, mark size=3, line width=1.1, mark options={solid}]
table {%
-20 -21.192813873291
-10 -20.5314464569092
-5 -19.2661781311035
0 -16.8337135314941
5 -13.4013061523438
};
\addlegendentry{DANSE - Mismtached}
\addplot [thick, orange, dashdotted, mark=square, mark size=3, line width=1.1,  mark options={solid}]
table {%
-20 -19.5719089508057
-10 -19.3565044403076
-5 -18.9108390808105
0 -17.7717819213867
5 -14.9769325256348
};
\addlegendentry{KalmanNet - Mismatched}
\end{axis}

\end{tikzpicture}}
    \caption{{Performance in case of mismatched process noise for the Lorenz attractor. At a fixed $\text{SMNR}=10$ dB, DANSE and KalmanNet are trained using a dataset $\mathcal{D}$ with $N=1000,T=100, \sigma_e^2=-10$ dB, and then tested at varying $\sigma_e^2$.}}
    \label{fig:nmse_lorenz_mismatched_process_noise}
\end{figure}
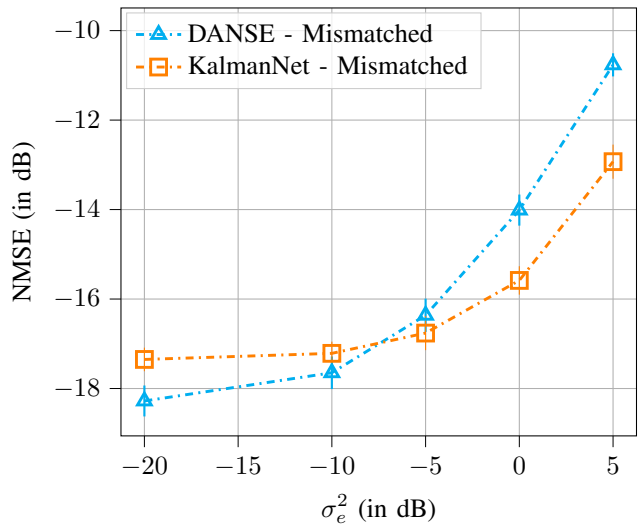

Here we study two cases as follows: 
\paragraph{Mismatched process} We study the robustness of {two} training-based methods - DANSE and KalmanNet - for a mismatched process. We generate a mismatched process by varying the process noise $\sigma_e^2$. A change in the process noise reflects a drift in the original SSM. Keeping SMNR = 10 dB, we train DANSE and KalmanNet using $\sigma_e^2=-10$ \text{dB} and then test at varying $\sigma_e^2$. An increase in $\sigma_e^2$ amounts to an increase in uncertainty (or randomness) in Lorenz SSM. The performances are shown in Fig. \ref{fig:nmse_lorenz_mismatched_process_noise}. We observe that the performances of both DANSE and KalmanNet deteriorate with increasing $\sigma_e^2$.
\paragraph{Mismatched measurement noise}
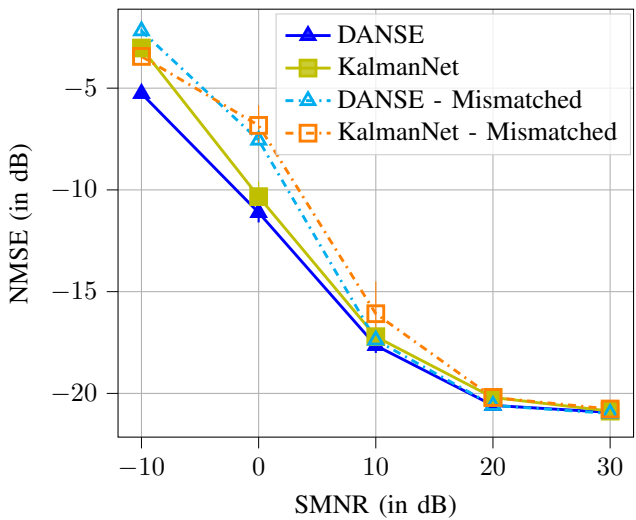
\begin{figure}[htbp]
    \centering
    \scalebox{1.0}{
\begin{tikzpicture}

\definecolor{darkgray176}{RGB}{176,176,176}
\definecolor{goldenrod1911910}{RGB}{191,191,0}
\definecolor{lightgray204}{RGB}{204,204,204}

\begin{axis}[
legend cell align={left},
legend style={fill opacity=0.8, draw opacity=1, text opacity=1, draw=lightgray204},
tick align=outside,
tick pos=left,
x grid style={darkgray176},
xlabel={SMNR (in dB)},
xmajorgrids,
xmin=-12, xmax=32,
xtick style={color=black},
y grid style={darkgray176},
ylabel={NMSE (in dB)},
ymajorgrids,
ymin=-35.9081557065249, ymax=-0.121219012141228,
ytick style={color=black}
]
\path [draw=blue, thick]
(axis cs:-10,-5.19774712622166)
--(axis cs:-10,-4.81431113183498);

\path [draw=blue, thick]
(axis cs:0,-13.1791091561317)
--(axis cs:0,-12.1042599081993);

\path [draw=blue, thick]
(axis cs:10,-20.8487460613251)
--(axis cs:10,-20.2165615558624);

\path [draw=blue, thick]
(axis cs:20,-27.733029127121)
--(axis cs:20,-27.2643845081329);

\path [draw=blue, thick]
(axis cs:30,-34.281476765871)
--(axis cs:30,-33.9819357097149);

\path [draw=goldenrod1911910]
(axis cs:-10,-3.40908762812614)
--(axis cs:-10,-2.77661564946175);

\path [draw=goldenrod1911910]
(axis cs:0,-11.5701081156731)
--(axis cs:0,-10.2161238789558);

\path [draw=goldenrod1911910]
(axis cs:10,-19.5968619138002)
--(axis cs:10,-19.1167611330748);

\path [draw=goldenrod1911910]
(axis cs:20,-27.3847634643316)
--(axis cs:20,-27.0142691284418);

\path [draw=goldenrod1911910]
(axis cs:30,-33.6552928686142)
--(axis cs:30,-33.4148670434952);

\path [draw=blue, thick]
(axis cs:-10,-1.94169250130653)
--(axis cs:-10,-1.74789795279503);

\path [draw=blue, thick]
(axis cs:0,-8.27599754929543)
--(axis cs:0,-7.57670173048973);

\path [draw=blue, thick]
(axis cs:10,-20.8481353521347)
--(axis cs:10,-20.2191177606583);

\path [draw=blue, thick]
(axis cs:20,-24.0520069301128)
--(axis cs:20,-23.536139279604);

\path [draw=blue, thick]
(axis cs:30,-31.6159167736769)
--(axis cs:30,-31.3511013537645);

\path [draw=goldenrod1911910]
(axis cs:-10,-4.7697996199131)
--(axis cs:-10,-3.91132602095604);

\path [draw=goldenrod1911910]
(axis cs:0,-11.3087015151978)
--(axis cs:0,-9.5156888961792);

\path [draw=goldenrod1911910]
(axis cs:10,-19.5969237536192)
--(axis cs:10,-19.1170769482851);

\path [draw=goldenrod1911910]
(axis cs:20,-24.1609058380127)
--(axis cs:20,-23.9309482574463);

\path [draw=goldenrod1911910]
(axis cs:30,-26.6053606420755)
--(axis cs:30,-26.4721845239401);

\addplot [thick, blue, mark=triangle*, mark size=3, line width=1.1, mark options={solid}]
table {%
-10 -5.00602912902832
0 -12.6416845321655
10 -20.5326538085938
20 -27.498706817627
30 -34.131706237793
};
\addlegendentry{DANSE}
\addplot [thick, goldenrod1911910, mark=square*, mark size=3, line width=1.1, mark options={solid}]
table {%
-10 -3.09285163879395
0 -10.8931159973145
10 -19.3568115234375
20 -27.1995162963867
30 -33.5350799560547
};
\addlegendentry{KalmanNet}
\addplot [thick, cyan, dashdotted, mark=triangle, mark size=3, line width=1.1, mark options={solid}]
table {%
-10 -1.84479522705078
0 -7.92634963989258
10 -20.5336265563965
20 -23.7940731048584
30 -31.4835090637207
};
\addlegendentry{DANSE-Mismatched}
\addplot [thick, orange, dashdotted, mark=square, mark size=3, line width=1.1,  mark options={solid}]
table {%
-10 -4.34056282043457
0 -10.4121952056885
10 -19.3570003509521
20 -24.0459270477295
30 -26.5387725830078
};
\addlegendentry{KalmanNet-Mismatched}
\end{axis}

\end{tikzpicture}}
    \caption{{Performance in case of mismatched measurement noise for the Lorenz attractor SSM. The `DANSE - Mismatched' and `KalmanNet - Mismatched' were trained at $\text{SMNR}=10$ dB and $\sigma_e^2=-10$ dB using a dataset $\mathcal{D}$ with $N=1000, T=100$, and then tested at varying $\text{SMNR}$.}}
    \label{fig:nmse_lorenz_mismatched_meas_noise}
\end{figure}
Now, we investigate the effect of a mismatch in the additive measurement noise $\mathbf{w}_t$ on the performances of DANSE and KalmanNet. The change in the measurement noise during testing reflects a change in the measurement setup. Keeping $\sigma_e^2$ fixed corresponding to $-10$ \text{dB}, we train DANSE and KalmanNet using $\text{SMNR} = 10$ dB and then test at varying SMNR. The performances are shown in Fig. \ref{fig:nmse_lorenz_mismatched_meas_noise}. We can observe that DANSE shows a degradation in performance at mismatched SMNR (i.e. except at $\text{SMNR} = 10$ dB that corresponds to the matched training condition). {DANSE appears to be more susceptible than KalmanNet to mismatched measurement noise at lower SMNR values, but less susceptible at higher SMNR values}. 
\paragraph{Remark} The susceptibility of DANSE to mismatched training and testing conditions is perhaps due to its lack of a-priori knowledge about the process. An educated guess to address the susceptibility is via a standard approach of multi-condition training, for example, if we train DANSE using training data comprised of several SMNRs and/or several process noise strengths. We do not further investigate the design of robust DANSE in this article. The design of robust DANSE using multi-condition training remains as part of future work.
\begin{figure*}[t]
    \centering
    \includegraphics[width=0.95\textwidth]{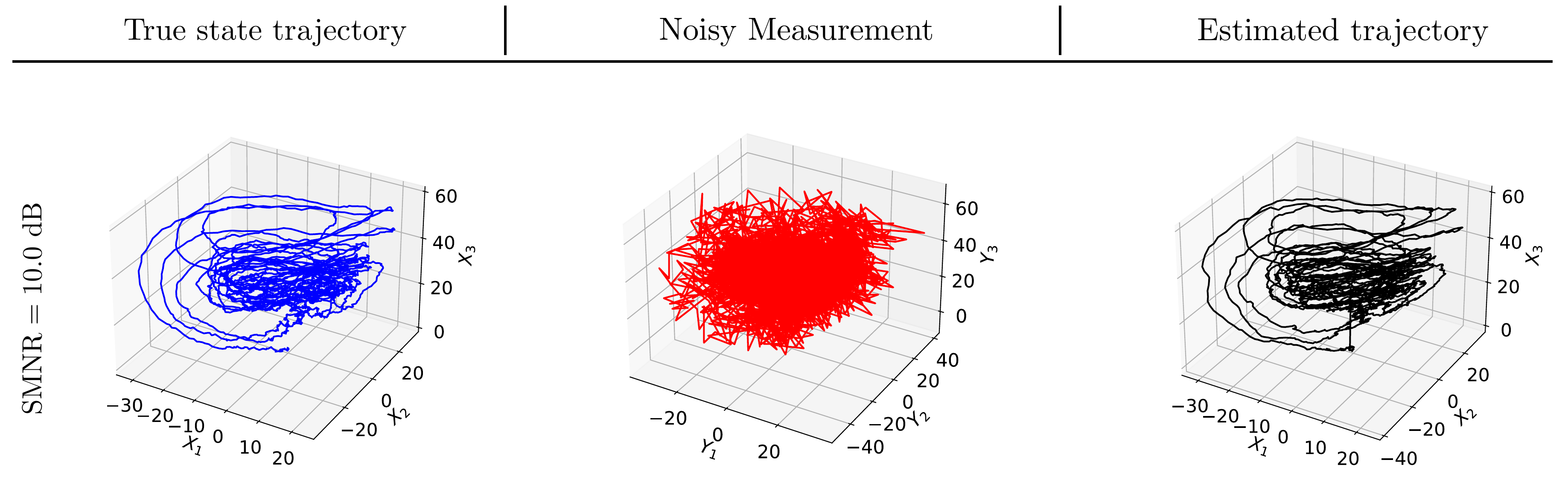}
    \caption{{State estimation for the Chen attractor using DANSE, at 10 dB SMNR with $\mathbf{H}_t = \mathbf{I}_3$. The true state trajectory at $\sigma_e^2 = -10$ dB is shown on the left, noisy measurement trajectory at 10 dB SMNR in the middle, and the estimated state trajectory on the right.}}
    \label{fig:chen_demonstrator}
\end{figure*}
\subsection{A limitation of DANSE (for subsampled measurements)}
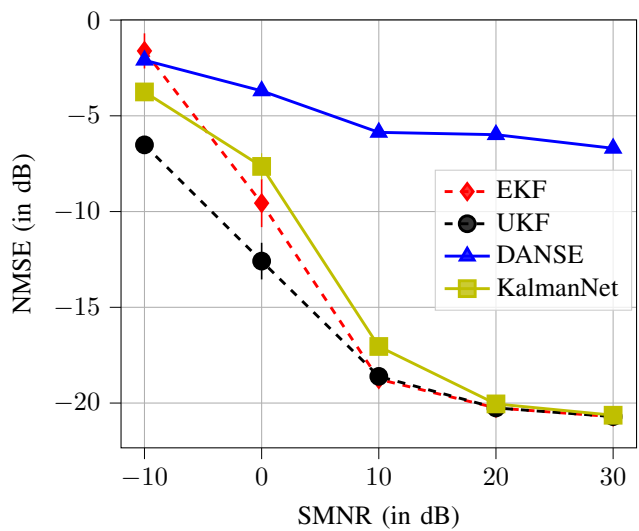
\begin{figure}[t]
    \centering
    \scalebox{1.0}{
\begin{tikzpicture}

\definecolor{darkgray176}{RGB}{176,176,176}
\definecolor{goldenrod1911910}{RGB}{191,191,0}
\definecolor{green01270}{RGB}{0,127,0}
\definecolor{lightgray204}{RGB}{204,204,204}

\begin{axis}[
legend cell align={left},
legend style={
  fill opacity=0.8,
  draw opacity=1,
  text opacity=1,
  at={(0.8,0.65)},
  anchor=north,
  draw=lightgray204
},
tick align=outside,
tick pos=left,
x grid style={darkgray176},
xlabel={SMNR (in dB)},
xmajorgrids,
xmin=-12, xmax=32,
xtick style={color=black},
y grid style={darkgray176},
ylabel={NMSE (in dB)},
ymajorgrids,
ymin=-33.7808993406594, ymax=1.58394868150353,
ytick style={color=black}
]

\path [draw=red, semithick]
(axis cs:-10,-2.35096061229706)
--(axis cs:-10,-1.02035820484161);

\path [draw=red, semithick]
(axis cs:0,-11.4245303869247)
--(axis cs:0,-8.72598421573639);

\path [draw=red, semithick]
(axis cs:10,-22.7821714580059)
--(axis cs:10,-21.9843461811543);

\path [draw=red, semithick]
(axis cs:20,-28.2248374521732)
--(axis cs:20,-27.734696239233);

\path [draw=red, semithick]
(axis cs:30,-31.8552244305611)
--(axis cs:30,-31.4143068194389);

\path [draw=black, semithick]
(axis cs:-10,-6.77563843131065)
--(axis cs:-10,-6.27034774422646);

\path [draw=black, semithick]
(axis cs:0,-14.668372631073)
--(axis cs:0,-12.6036496162415);

\path [draw=black, semithick]
(axis cs:10,-22.7667089402676)
--(axis cs:10,-22.0041307508945);

\path [draw=black, semithick]
(axis cs:20,-28.1581853330135)
--(axis cs:20,-27.6823298037052);

\path [draw=black, semithick]
(axis cs:30,-31.723675146699)
--(axis cs:30,-31.2769008725882);

\path [draw=blue, thick]
(axis cs:-10,-2.20523678511381)
--(axis cs:-10,-2.08008731156588);

\path [draw=blue, thick]
(axis cs:0,-3.83202613145113)
--(axis cs:0,-3.69235359877348);

\path [draw=blue, thick]
(axis cs:10,-6.03037321567535)
--(axis cs:10,-5.91142594814301);

\path [draw=blue, thick]
(axis cs:20,-6.10774786025286)
--(axis cs:20,-5.98525827378035);

\path [draw=blue, thick]
(axis cs:30,-6.82546693086624)
--(axis cs:30,-6.69071501493454);

\path [draw=goldenrod1911910]
(axis cs:-10,-4.10846716165543)
--(axis cs:-10,-3.40467065572739);

\path [draw=goldenrod1911910]
(axis cs:0,-8.87623709440231)
--(axis cs:0,-7.35528832674026);

\path [draw=goldenrod1911910]
(axis cs:10,-19.3684912621975)
--(axis cs:10,-18.4974221289158);

\path [draw=goldenrod1911910]
(axis cs:20,-27.0873891711235)
--(axis cs:20,-25.9958208203316);

\path [draw=goldenrod1911910]
(axis cs:30,-30.6515449285507)
--(axis cs:30,-30.1467720270157);

\addplot [semithick, red, dashed, mark=diamond*, mark size=3, line width=1.1, mark options={solid}]
table {%
-10 -1.68565940856934
0 -10.0752573013306
10 -22.3832588195801
20 -27.9797668457031
30 -31.634765625
};
\addlegendentry{EKF}
\addplot [semithick, dashed, black, mark=*, mark size=3, line width=1.1, mark options={solid}]
table {%
-10 -6.52299308776855
0 -13.6360111236572
10 -22.3854198455811
20 -27.9202575683594
30 -31.5002880096436
};
\addlegendentry{UKF}
\addplot [thick, blue, mark=triangle*, mark size=3, line width=1.1, mark options={solid}]
table {%
-10 -2.14266204833984
0 -3.7621898651123
10 -5.97089958190918
20 -6.0465030670166
30 -6.75809097290039
};
\addlegendentry{DANSE}
\addplot [goldenrod1911910, mark=square*, mark size=3, line width=1.1, mark options={solid}]
table {%
-10 -3.75656890869141
0 -8.11576271057129
10 -18.9329566955566
20 -26.5416049957275
30 -30.3991584777832
};
\addlegendentry{KalmanNet}
\end{axis}

\end{tikzpicture}}
    \caption{{Effect of subsampled measurements using $\mathbf{H}_t = \mathbf{H}$ as a $2 \times 3$ Gaussian random matrix. DANSE and KalmanNet were both trained using a dataset $\mathcal{D}$ with $N=1000,T=100$. DANSE shows a limitation.}}
    \label{fig:subsampled_rn2}
\end{figure}

While we {have} developed DANSE in an unsupervised manner and {have} shown its competitive performance against several methods, a question can be as follows: is there any limitation of DANSE? Is there a scenario where DANSE has a high limitation, but the other competitors do well? Here we demonstrate such a scenario experimentally. We perform an experiment for subsampled measurements in the case of Lorenz attractor SSM. We implement this using a random measurement matrix $\mathbf{H}_t = \mathbf{H}$ of size $2 \times 3$ with i.i.d. entries sampled from $\mathcal{N}\left(0, 1\right)$. This is a case of subsampling with $\frac{2}{3}$ ratio ($\frac{n}{m} = \frac{2}{3}$). Note that this is an under-determined estimation and learning problem. We then generate training data using this random matrix $\mathbf{H}$ and train DANSE appropriately. The performance is shown in Fig. \ref{fig:subsampled_rn2}, where DANSE fails to perform. On the other hand, EKF, UKF, and KalmanNet do well. Our guess is that DANSE suffers since it has no a-priori knowledge of the process. Therefore, it lacks appropriate regularization for solving the under-determined estimation problem. We do not further investigate the design of DANSE for subsampled measurements in this article. This remains a part of future work.


\begin{figure*}[t]
    \centering
    \includegraphics[width=0.95\textwidth]{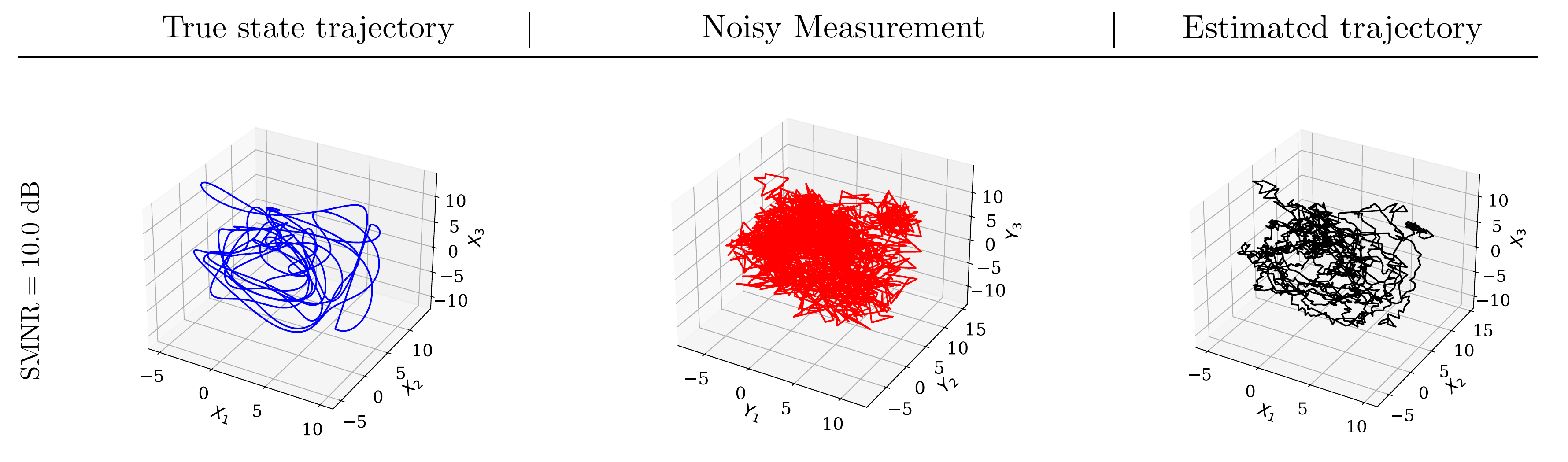}
    \caption{{State estimation for the 20-dimensional Lorenz-96 attractor using DANSE, at 10 dB SMNR with $\mathbf{H}_t = \mathbf{I}_{20}$. A true state trajectory of the first 3 out of 20 states at $\sigma_e^2 = -10$ dB is shown on the left, the corresponding noisy measurement trajectory at 10 dB SMNR in the middle, and the corresponding estimated state trajectory using DANSE on the right.}}
    \label{fig:lorenz96_demonstrator}
\end{figure*}
\subsection{Experiments for another nonlinear SSM - Chen attractor}\label{sec:chen_experiments}
So far we have shown results for one nonlinear SSM - the Lorenz attractor SSM. A natural question is how DANSE performs for other nonlinear SSMs. Or a non-trivial question can be: what will be the complexity level of a nonlinear SSM such that DANSE can show a good performance? As we do not have any theoretical arguments at this point to answer the questions, we conduct experiments for another chaotic attractor, called Chen attractor \cite{chen1999yet, vcelikovsky2005generalized}. The experiment is to demonstrate the behavior of DANSE in the case of another nonlinear SSM. 
\begin{figure*}[t]
\begin{minipage}{0.31\textwidth}
    \centering
    \scalebox{1.0}{
\begin{tikzpicture}

\definecolor{darkgray176}{RGB}{176,176,176}

\begin{axis}[
tick align=outside,
tick pos=left,
x grid style={darkgray176},
xmin=-0.5, xmax=1999.5,
xtick style={color=black},
y grid style={darkgray176},
ymin=-0.5, ymax=19.5,
ytick style={color=black},
width=\textwidth,
height=0.9\textwidth
]
\addplot graphics [includegraphics cmd=\pgfimage,xmin=-0.5, xmax=1999.5, ymin=-0.5, ymax=19.5] {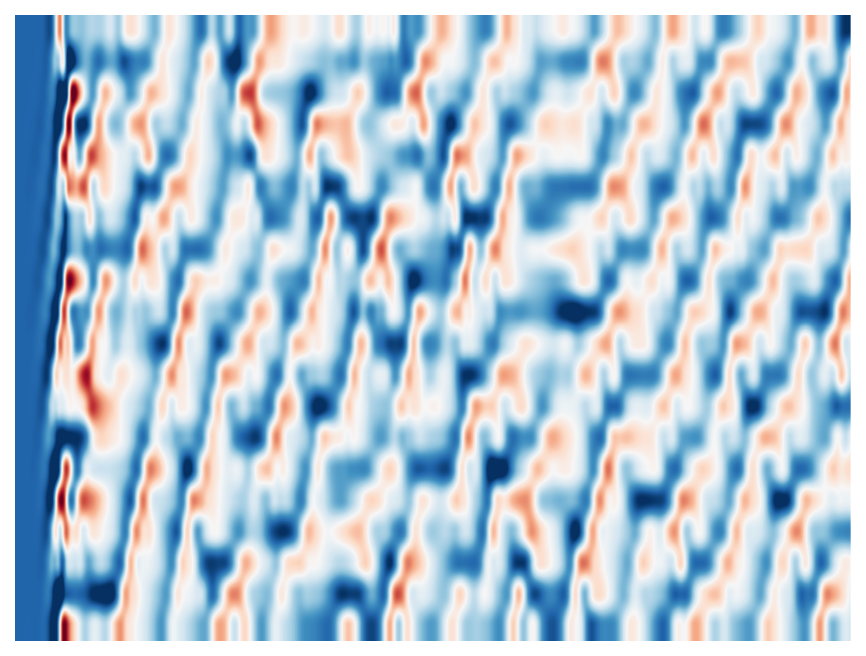};
\end{axis}

\end{tikzpicture}}
    \subcaption{}
\end{minipage}  
\begin{minipage}{0.31\textwidth}
    \centering
    \scalebox{1.0}{
\begin{tikzpicture}

\definecolor{darkgray176}{RGB}{176,176,176}

\begin{axis}[
tick align=outside,
tick pos=left,
x grid style={darkgray176},
xmin=-0.5, xmax=1999.5,
xtick style={color=black},
y grid style={darkgray176},
ymin=-0.5, ymax=19.5,
ytick style={color=black},
width=\textwidth,
height=0.9\textwidth
]
\addplot graphics [includegraphics cmd=\pgfimage,xmin=-0.5, xmax=1999.5, ymin=-0.5, ymax=19.5] {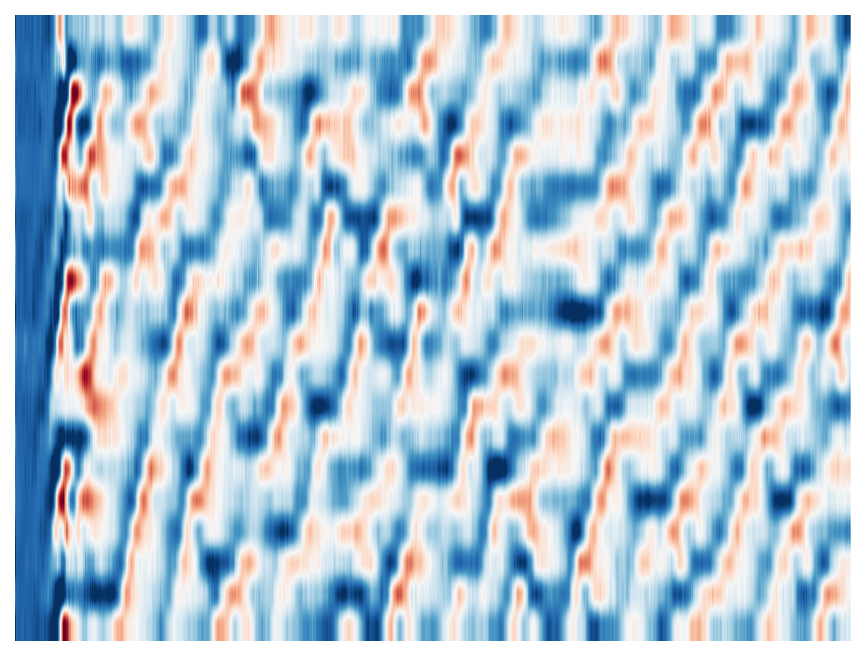};
\end{axis}

\end{tikzpicture}}
    \subcaption{}
\end{minipage}
\begin{minipage}{0.31\textwidth}
    \centering
    \scalebox{1.0}{
\begin{tikzpicture}

\definecolor{darkgray176}{RGB}{176,176,176}

\begin{axis}[
colorbar,
colorbar style={ylabel={}},
colormap={mymap}{[1pt]
  rgb(0pt)=(0.403921568627451,0,0.12156862745098);
  rgb(1pt)=(0.698039215686274,0.0941176470588235,0.168627450980392);
  rgb(2pt)=(0.83921568627451,0.376470588235294,0.301960784313725);
  rgb(3pt)=(0.956862745098039,0.647058823529412,0.509803921568627);
  rgb(4pt)=(0.992156862745098,0.858823529411765,0.780392156862745);
  rgb(5pt)=(0.968627450980392,0.968627450980392,0.968627450980392);
  rgb(6pt)=(0.819607843137255,0.898039215686275,0.941176470588235);
  rgb(7pt)=(0.572549019607843,0.772549019607843,0.870588235294118);
  rgb(8pt)=(0.262745098039216,0.576470588235294,0.764705882352941);
  rgb(9pt)=(0.129411764705882,0.4,0.674509803921569);
  rgb(10pt)=(0.0196078431372549,0.188235294117647,0.380392156862745)
},
point meta max=10,
point meta min=-10,
tick align=outside,
tick pos=left,
x grid style={darkgray176},
xmin=-0.5, xmax=1999.5,
xtick style={color=black},
y grid style={darkgray176},
ymin=-0.5, ymax=19.5,
ytick style={color=black},
width=\textwidth,
height=0.9\textwidth
]
\addplot graphics [includegraphics cmd=\pgfimage,xmin=-0.5, xmax=1999.5, ymin=-0.5, ymax=19.5] {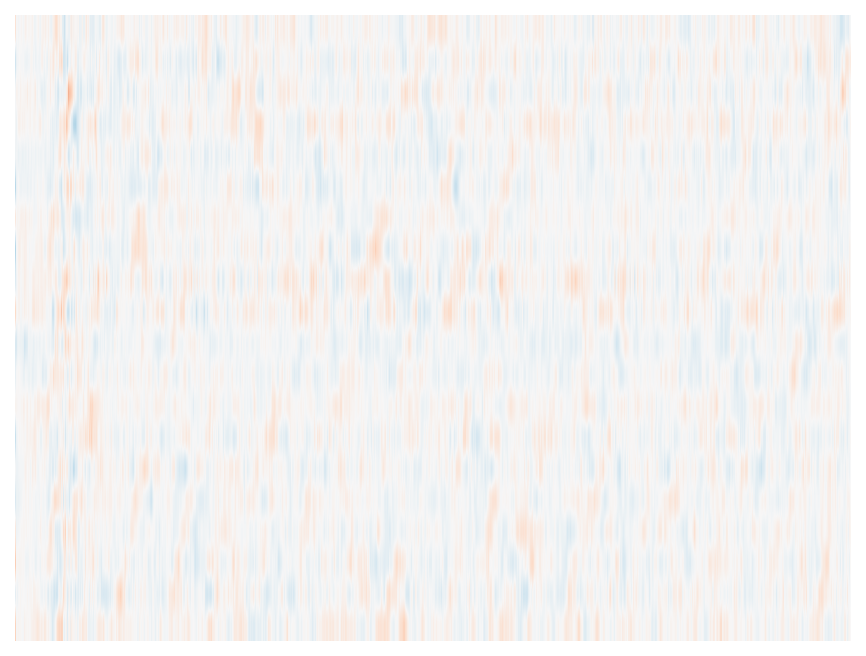};
\end{axis}

\end{tikzpicture}}
    \subcaption{}
\end{minipage}
\caption{
{Image plot for state estimation in case of a 20-dimensional Lorenz-96 attractor. Fig. (a) shows the image plot of a 20-dimensional true state trajectory of length $T_{\text{test}}=2000$, (b) shows the image plot for the corresponding state estimate using DANSE, and finally, (c) shows the image plot of the error between the true state and the estimated state trajectory using DANSE.}}
\label{fig:lorenz96}
\end{figure*} 
For the Chen attractor SSM, we performed experiments similar to the Lorenz attractor SSM in section \ref{sec:Lorenz_SSM}. Instead of reporting several results, we show DANSE performance visually in Fig. \ref{fig:chen_demonstrator} for one randomly chosen trajectory at $\text{SMNR} = 10$ dB. The figure helps to visualize the state trajectory for the Chen attractor SSM, the corresponding measurement trajectory at $10$ dB SMNR, and finally the estimated trajectory using DANSE. In this experiment, the NMSE of DANSE is found to be $-22.40 \pm 0.29$ \text{dB}, while for EKF and UKF the NMSE values are roughly the same and equal to $-22.73 \pm 0.27$ \text{dB} {for a test dataset having $N_{\text{test}} = 100, T_{\text{test}} = 5000$. The slightly higher value of $T_{\text{test}}$ compared to the Lorenz attractor is owing to the use of a smaller discretization step size for avoiding numerical instability.}  Hence, for the Chen attractor SSM, DANSE also shows competitive performance. The Chen attractor model that we use for our simulation study is shown in appendix \ref{sec:chen}.

\subsection{{Experiments for a high-dimensional SSM - Lorenz-96 attractor}}\label{sec:lorenz_96_experiments}

{
In this subsection, we demonstrate the performance of DANSE for high-dimensional state estimation, based on  Lorenz-96 attractor SSM \cite{lorenz1996predictability, lorenz1998optimal}. The Lorenz-96 attractor model that we use for our simulation study is shown in appendix \ref{sec:lorenz96}. We use 20-dimensional SSM for the experiments. Given the functional form in \eqref{eq:lorenz96_state}, we remark that implementing an EKF or UKF in this scenario is non-trivial. This is due to linearization difficulties in the case of discretizing ODEs. 
However, DANSE doesn't require explicit knowledge of the underlying process and thus offers a modeling advantage. In this experiment, the NMSE of DANSE is found to be $-17.01 \pm 0.16$ \text{dB} for a state estimation task using $\text{SMNR} = 10 \,\, \text{dB}$ for a test dataset having $N_{\text{test}} = 100, T_{\text{test}} = 2000$. We visually demonstrate the performance of DANSE on the Lorenz-96 attractor in Figs. \ref{fig:lorenz96_demonstrator}, \ref{fig:lorenz96} for a randomly selected test-set trajectory. Fig. \ref{fig:lorenz96_demonstrator} to visualize the state trajectory for the Chen attractor SSM, the corresponding measurement trajectory at $10$ dB SMNR, and finally the estimated trajectory using DANSE for the first 3 out of 20 states. Since visualizing all the 20 states in a time plot is cumbersome, we take inspiration from the work in \cite{pawar2020long}, and display an image plot of the 20-dimensional state trajectory versus time and its corresponding estimate using DANSE in Fig. \ref{fig:lorenz96}.
}
   
\section{Conclusions and future scopes} \label{sec:conclusions}
Experimental results show that the proposed DANSE is competitive {and works {also} for high-dimensional state estimation}. The development of DANSE concludes that it is possible to design an unsupervised learning-based data-driven method to estimate the non-linear state of a model-free process using a training dataset of linear measurements. In addition, a major conclusion is that the amount of training data is a key factor contributing to good performance. The NMSE performance improvement with the size of training data follows a power law behavior. It also shows that unsupervised learning faces difficulty in handling an under-determined system for a model-free process (the sub-sampled measurement system that we investigated).

Several questions remain open. (a) How to handle nonlinear measurements instead of linear measurements? (b) How to handle unknown Gaussian measurement noise statistics, for example, the unknown covariance matrix? (c) How to handle non-Gaussian measurement noise statistics? (d) How to handle mismatched training-and-testing conditions for robustness? (e) How to handle data-limited scenarios, for example, a limited amount of training data, and/or sub-sampled measurements? We hope that the above questions will pose new research challenges in the future. 

\appendix
\subsection{Chen attractor model} \label{sec:chen}
As in the case of the Lorenz attractor SSM in section \ref{sec:Lorenz_SSM}, the Chen attractor SSM \cite{chen1999yet} is in a discretized form as 
\begin{equation}\label{eq:chen_state}
    \mbx_{t}=\mathbf{F}^{\prime}_t(\mbx_{t-1})\mbx_{t-1} + \mathbf{e}_t \in \mathbb{R}^3,
\end{equation}
where $\mathbf{e}_t\sim\mathcal{N}(\boldsymbol{0},\mathbf{C}_e)$ with $\mathbf{C}_e=\sigma_e^2 \mathbf{I}_{3}$, and
\begin{eqnarray}\label{eq:F_chen}
\mathbf{F}^{\prime}_t(\mbx_{t-1}) = \exp\left(\begin{bmatrix}
-35 & 35 & 0 \\
-7 & 28 & -x_{t-1, 1} \\
0 & x_{t-1, 1} & -{3} \\ 
\end{bmatrix}\Delta^{\prime} \right),
\end{eqnarray}
with the step-size $\Delta^{\prime}=0.002 \text{ seconds}$. The discretization is performed similarly as in \cite{garcia2019combining}. We note that in the case of the Chen attractor \cite{chen1999yet}, we require a smaller step size $\Delta^{\prime}$, to avoid instability during simulation. In our simulations, we use a finite-Taylor series approximation of $5'\text{th}$ order for \eqref{eq:F_chen}. The measurement system is linear, where $\mathbf{H}_t=\mathbf{I}_3$ in \eqref{eq:measurement}. The measurement noise is $\mathbf{w}_t\sim\mathcal{N}(\boldsymbol{0},\mathbf{C}_w)$ with $\mathbf{C}_w=\sigma_w^2 \mathbf{I}_{3}$, while $\sigma_e^2$ corresponds to $-10$\text{ dB}. 
\subsection{{High dimensional Lorenz-96 attractor model}} \label{sec:lorenz96}
{
Unlike the Lorenz SSM in \eqref{eq:lorenz_state}, the Lorenz-96 SSM cannot be written very easily in a matrix form, and is usually expressed in a continuous format, with the $j'\text{th}$-coordinate being shown as
\begin{equation}\label{eq:lorenz96_state}
    \frac{d{x}_{{t}_c,j}}{dt_c}=\left(x_{{t}_c,j+1} - x_{{t}_c,j-2}\right)x_{{t}_c,j-1} - x_{{t}_c,j} + {F}_{{t}_c} \in \mathbb{R},
\end{equation}
where $j=1, 2, \ldots, m$, with the conventions that $x_{{t}_c,0} = x_{{t}_c,m}, x_{{t}_c,-1} = x_{{t}_c,m-1}, x_{{t}_c,m+1} = x_{1}$. With a slight abuse of notation, ${t}_c$ denotes the continuous time index in this case. We model ${F}_{t_c}\sim\mathcal{N}(\mu_{F}, {\sigma}^{2}_{F})$ as a time-varying process-noise with mean $\mu_{F} = 8$ with ${\sigma}^{2}_{F}$ corresponding to $-10.0$ dB. The differential equation in \eqref{eq:lorenz96_state} is simulated using a $4'\text{th}$ order Runge-Kutta method, with the step-size $\Delta = 0.01 \text{ seconds}$. The simulation in \eqref{eq:lorenz96_state} is carried out for $m=20$ such states $\lbrace x_{t_c, j} \rbrace_{j=1}^{20} \bigg\vert_{t_c = t\Delta}$ which gives the discrete-time state vector $\mathbf{x}_t = \left[ x_{t,1}, \ldots, x_{t,20} \right]^{\top}$. The measurement system is linear, where $\mathbf{H}_t=\mathbf{I}_{20}$ in  \eqref{eq:measurement}. The measurement noise is $\mathbf{w}_t\sim\mathcal{N}(\boldsymbol{0},\mathbf{C}_w)$ with $\mathbf{C}_w=\sigma_w^2 \mathbf{I}_{20}$. 
}
{In our experiments we used $\text{SMNR} = 10.0 \text{ dB}$.}
\bibliographystyle{IEEEbib}
\bibliography{refs}
\end{document}